\documentclass[useAMS,usenatbib]{mn2e}
\usepackage{float}
\usepackage{caption}
\usepackage{graphicx}
\usepackage{epstopdf}
\usepackage{amsmath}
\usepackage{placeins}
\title[The old nuclear star cluster in the Milky Way]{The old nuclear
  star cluster in the Milky Way: dynamics, mass, statistical parallax,
  and black hole mass} \author[S. Chatzopoulos et
al.]{S. Chatzopoulos$^{1}$\thanks{E-mail: sotiris@mpe.mpg.de,
    gerhard@mpe.mpg.de}, T. K. Fritz$^{2}$, O. Gerhard$^{1}$,
  S. Gillessen$^{1}$, C. Wegg$^{1}$, R. Genzel$^{1,3}$, \and
  O. Pfuhl$^{1}$
  \\
  $^{1}$Max Planck Institut fur Extraterrestrische Physic, Postfach
  1312, D-85741, Garching Germany
  \\
  $^{2}$Department of Astronomy, University of Virginia, 530 McCormick Road Charlottesville VA 22904-4325 USA
  \\
  $^{3}$Department of Physics, Le Conte Hall, University of California, 94720 Berkeley, USA}

\begin{document}

\date{Submitted 2014 March 15}

\pagerange{\pageref{firstpage}--\pageref{lastpage}} \pubyear{2002}

\maketitle

\label{firstpage}

\begin{abstract}
  We derive new constraints on the mass, rotation, orbit structure and
  statistical parallax of the Galactic old nuclear star cluster and
  the mass of the supermassive black hole.  We combine star counts and
  kinematic data from \citet{fc2014}, including 2'500 line-of-sight
  velocities and 10'000 proper motions obtained with VLT instruments.
  We show that the difference between the proper motion dispersions
  $\sigma_l$ and $\sigma_b$ cannot be explained by rotation, but is a
  consequence of the flattening of the nuclear cluster. We fit the
  surface density distribution of stars in the central $1000''$ by a
  superposition of a spheroidal cluster with scale $\sim 100''$ and a
  much larger nuclear disk component. We compute the self-consistent
  two-integral distribution function $f(E,L_z)$ for this density
  model, and add rotation self-consistently.  We find that: (i) The
  orbit structure of the $f(E,L_z)$ gives an excellent match to the
  observed velocity dispersion profiles as well as the proper motion
  and line-of-sight velocity histograms, including the double-peak in
  the $v_l$-histograms. (ii) This requires an axial ratio near
  $q_1=0.7$ consistent with our determination from star counts, $q_1 = 0.73
  \pm 0.04$ for $r<70''$. (iii) The nuclear star cluster is
  approximately described by an isotropic rotator model. (iv) Using
  the corresponding Jeans equations to fit the proper motion and
  line-of-sight velocity dispersions, we obtain best estimates for the
  nuclear star cluster mass, black hole mass, and distance
  ${M_*}(r\!<\!100'')\!=\!(8.94\!\pm\! 0.31{|_{\rm stat}}
  \!\pm\!0.9{|_{\rm syst}})\!\times\! {10^6}{M_\odot}$, ${M_\bullet }
  \!=\! (3.86\!\pm\!0.14{|_{\rm stat} \!\pm\! 0.4{|_{\rm syst}}})
  \!\times\! {10^6}{M_\odot }$, and ${R_0} \!=\! 8.27 \!\pm\!
  0.09{|_{\rm stat}}\!\pm\! 0.1{|_{\rm syst}}$ kpc, where the
    estimated systematic errors account for additional uncertainties
    in the dynamical modeling. (v) The combination of the cluster
  dynamics with the S-star orbits around Sgr A$^*$ strongly reduces
  the degeneracy between black hole mass and Galactic centre distance
  present in previous S-star studies.  A joint statistical analysis
  with the results of \citet{ge2009} gives ${M_\bullet } \!=\!
  (4.23\!\pm\!0.14)\!\times\! {10^6}{M_\odot}$ and ${R_0} \!=\! 8.33
  \!\pm\! 0.11$ kpc.
\end{abstract}


\section{INTRODUCTION}

Nuclear star clusters (NSC) are located at the centers of most
spiral galaxies \citep{carollo1997,boeker2002}. They are more
luminous than globular clusters \citep{boeker2004}, have masses of
order $\sim10^6-10^7 M_\odot$ \citep{walcher2005}, have complex star
formation histories \citep{rossa2006,seth2006}, and obey
scaling-relations with host galaxy properties as do central
supermassive black holes \citep[SMBH;][]{ferrarese2006,wehner2006};
see \citet{boeker2010} for a review. Many host an AGN, i.e., a SMBH
\citep{seth2008}, and the ratio of NSC to SMBH mass varies widely
\citep{graham2009, kormendy2013}.

The NSC of the Milky Way is of exceptional interest because of its
proximity, about 8 kpc from Earth. It extends up to several hundred
arcsecs from the center of the Milky Way (Sgr A*) and its mass within
1 pc is $\sim 10^6M_\odot$ with $\sim50\%$ uncertainty \citep{sm2009,geisen2010}.
There is strong evidence that the center of the NSC hosts a SMBH of
several million solar masses. Estimates from stellar orbits show that
the SMBH mass is ${M_\bullet } = (4.31 \pm 0.36)\times{10^6}{M_\odot}$
\citep{schoe2002,ghez2008,ge2009}. Due to its proximity, individual
stars can be resolved and number counts can be derived; however, due
to the strong interstellar extinction the stars can only be observed
in the infrared. A large number of proper motions and line-of-sight
velocities have been measured, and analyzed with spherical models to
attempt to constrain the NSC dynamics and mass
\citep{hr1996,gt1996,genz2000,tg2008,sm2009,fc2014}.
 
The relaxation time of the NSC within 1 pc is ${t_r} \sim{10^{10}}$ yr
\citep{a2005, m2013}, indicating that the NSC is not fully relaxed and
is likely to be evolving. One would expect from theoretical models
that, if relaxed, the stellar density near the SMBH should be
steeply-rising and form a \citet{bw1976} cusp. In contrast,
observations by \citet{dg2009,b2009,b2010} show that the distribution
of old stars near the SMBH appears to have a core.  Understanding the
nuclear star cluster dynamics may therefore give useful constraints on
the mechanisms by which it formed and evolved \citep{m2010}.

In this work we construct axisymmetric Jeans and two-integral
distribution function models based on stellar number counts, proper
motions, and line-of-sight velocities.  We describe the data briefly in
Section~\ref{sDataset}; for more detail the reader is referred to the
companion paper of \citet{fc2014}.  In Section~\ref{sSpherical} we
carry out a preliminary study of the NSC dynamics using isotropic
spherical models, in view of understanding the effect of 
rotation on the data. In Section \ref{sAxis} we describe our
axisymmetric models and show that they describe the kinematic
properties of the NSC exceptionally well. By applying a $\chi^2$
minimization algorithm, we estimate the mass of the cluster, the SMBH
mass, and the NSC distance. We discuss our results and summarize our
conclusions in Section~\ref{s_discussion}. The Appendix contains some
details on our use of the \cite{qh1995} algorithm to calculate the
two-integral distribution function for the fitted density model.

\section[]{DATASET}
\label{sDataset}

We first give a brief description of the data set used for our
dynamical analysis. These data are taken from \citet{fc2014} and are
thoroughly examined in that paper, which should be consulted for more
details.  The coordinate system used is a shifted Galactic coordinate
system ($l^*,b^*$) where Sgr A* is at the center and ($l^*,b^*$) are
parallel to Galactic coordinates ($l,b$). In the following we always
refer to the shifted coordinates but will omit the asterisks for
simplicity. The dataset consists of stellar number densities, proper
motions and line-of-sight velocities. We use the stellar number
density map rather than the surface brightness map because it is less
sensitive to individual bright stars and non-uniform extinction.

The stellar number density distribution is constructed from NACO
high-resolution images for $R_{\rm box}<20''$, in a similar way as in
\citet{schoedel2010}, from HST WFC3/IR data for $20''<R_{\rm box}<66''$,
and from VISTA-VVV data for $66''<R_{\rm box}<1000''$.

\begin{figure}
\centering
\includegraphics[width=\linewidth]{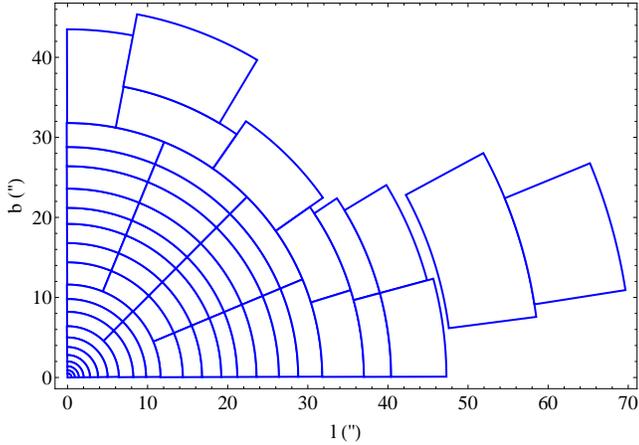}
\caption{Binning of the proper motion velocities. The stars are binned
  into cells according to their distance from Sgr A* and their smallest
  angle to the Galactic plane \citep{fc2014}.}
\label{plot_5}
\end{figure}

The kinematic data include proper motions for $\sim$10'000 stars
obtained from AO assisted images. The proper motion stars are binned
into 58 cells \citep[Figure \ref{plot_5};][]{fc2014} according to
distance from Sgr A* and the Galactic plane. This binning assumes that
the NSC is symmetric with respect to the Galactic plane and with
respect to the $b$-axis on the sky, consistent with axisymmetric
dynamical modeling. The sizes of the bins are chosen such that all
bins contain comparable numbers of stars, and the velocity
dispersion gradients are resolved, i.e., vary by less than the error
bars between adjacent bins. 

Relative to the large velocity dispersions at the Galactic
center (100 km/s), measurement errors for individual stars are
typically $\sim10\%$, much smaller than in typical globular cluster
proper motion data where they can be $\sim50\%$ (e.g., in Omega Cen;
\cite{v2006}). Therefore corrections for these measurement errors
are very small.

\begin{figure}
\centering
\includegraphics[width=\linewidth]{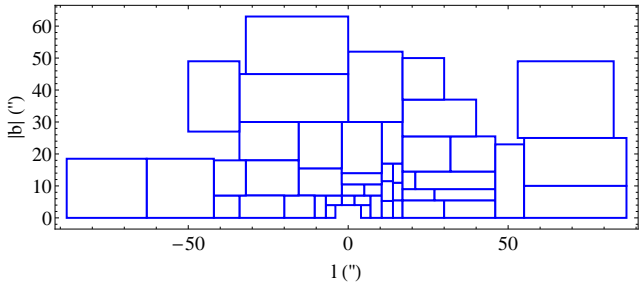}
\caption{Binning of the line-of-sight velocities. The stars are binned
  into 46 rectangular outer cells plus 6 rectangular rings at the
  center. The latter are located within the white area around $l\!=b\!=0$
  and are not shown in the plot; see App.~E of \citet{fc2014}.}
\label{plot_6}
\end{figure}

We also use $\sim$2'500 radial velocities obtained from SINFONI
integral field spectroscopy. The binning of the radial velocities is
shown in Fig.~\ref{plot_6}. There are 46 rectangular outer bins as
shown in Fig.~\ref{plot_6} plus 6 small rectangular rings around the
center \citep[not shown; see App.~E of][]{fc2014}. Again the outer
bins are chosen such that they contain similar numbers of stars
and the velocity dispersion gradients are resolved.  The
distribution of radial velocity stars on the sky is different from the
distribution of proper motion stars, and it is not symmetric with
respect to $l=0$.  Because of this and the observed rotation, the
binning is different, and extends to both positive and negative
longitudes. Both the proper motion and radial velocity binning are
also used in \citet{fc2014} and some tests are described in that
paper.

Finally, we compare our models with (but do not fit to) the
kinematics derived from about 200 maser velocities at $r > 100''$
\citep[from][]{lindquist1992,deguchi2004}. As for the proper motion
and radial velocity bins, we use the mean velocities and velocity
dispersions as derived in \citet{fc2014}.

The assumption that the NSC is symmetric with respect to the Galactic
plane and the $b=0$ axis is supported by the recent Spitzer/IRAC
photometry \citep{schodel2014} and by the distribution of proper
motions \citep{fc2014}. The radial velocity data at intermediate radii
instead show an apparent misalignment with respect to the
Galactic plane, by $\sim 10^{\circ}$; see \citet{feldmeier2014} and
\citet{fc2014}.  We show in Section~\ref{s_distance} that, even if
confirmed, such a misaligned structure would have minimal impact on
the results obtained here with the symmetrised analysis.

\section[]{SPHERICAL MODELS OF THE NSC}
\label{sSpherical}

In this section we study the NSC using the preliminary assumption that
the NSC can be described by an isotropic distribution function (DF) depending
only on energy. We use the DF to predict the kinematical data of the
cluster. Later we add rotation self-consistently to the model. The
advantages of using a distribution function instead of common Jeans
modeling are that (i) we can always check if a DF is positive and
therefore if the model is physical, and (ii) the DF provides us with
all the moments of the system. For the rest of the paper we use $(r,\theta ,\varphi )$
for spherical and $(R,\varphi ,z)$ for cylindrical coordinates, with $\theta=0$
corresponding to the z-axis normal to the equatorial plane of the NSC.

\begin{figure}
\centering
\includegraphics[width=\linewidth]{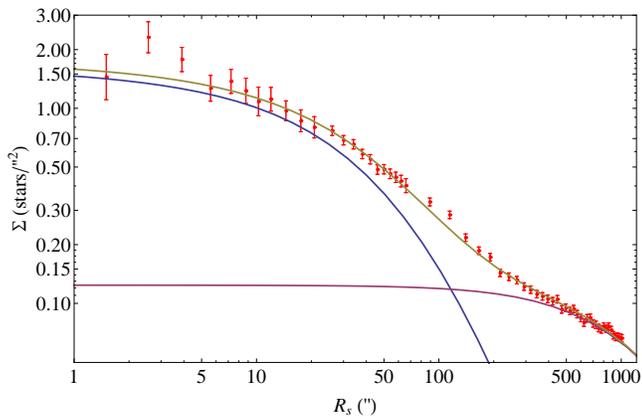}
\caption{A combination of two $\gamma$-models gives an accurate
  approximation to the spherically averaged number density of
  late-type stars versus radius on the sky (points with error bars).
  Blue line: inner component, purple line: outer component, brown
  line: both components.}
\label{plot_1}
\end{figure}

\subsection[]{Mass model for the NSC}
\label{oneIntegralDF}

The first step is to model the surface density. We use the well-known
one-parameter family of spherical $\gamma$-models \citep{d1993}:
\begin{align}
\rho_{\gamma} (r) = \frac{{3 - \gamma }}{{4\pi }}\frac{{M\,a }} 
   {{{r^\gamma }{{(r + a)}^{4 - \gamma }}}}\,,\,0 \le \gamma  < 3
\end{align}
where $a$ is the scaling radius and $M$ the total mass.The model
behaves as $\rho \sim {r^{ - \gamma }}$ for ${r \to 0}$ and $\rho \sim
{r^{ - 4}}$ for $r \to \infty $. Dehnen $\gamma$ models are equivalent
to the $\eta$-models of \cite{tr1994} under the transformation
$\gamma=3-\eta$. Special cases are the \cite{j1983} and \cite{h1990}
models for $\gamma=2$ and $\gamma=1$ respectively. For $\gamma=3/2$
the model approximates de Vaucouleurs $R^{1/4}$ law. In order to
improve the fitting of the surface density we use a combination of two
$\gamma$-models, i.e.
\begin{align}
\label{eqDenSph}
\rho(r) = \sum\limits_{i = 1}^2 {\frac{{3 - {\gamma_i}}}{{4\pi }} 
  \frac{{M_i\,a_i }}{{{r^{{\gamma_i}}}{{(r + a_i )}^{4 - {\gamma_i}}}}}}.
\end{align}
The use of a two-component model will prove convenient later when
we move to the axisymmetric case. The projected density is
\begin{align}
\label{eqSDensity}
\Sigma (R_s) = 2\int_{R_s}^\infty  {{\rho}(r)r} /{({r^2} - {R_s^2})^{1/2}}dr
\end{align}
and can be expressed in terms of elementary functions for integer
$\gamma$, or in terms of elliptic integrals for half-integer
$\gamma$. For arbitrary $\gamma_1$ and $\gamma_2$ the surface density
can only be calculated numerically using
equation~(\ref{eqSDensity}). The surface density diverges for
$\gamma>1$ but is finite for $\gamma<1$.

\begin{figure}
\centering
\includegraphics[width=\linewidth]{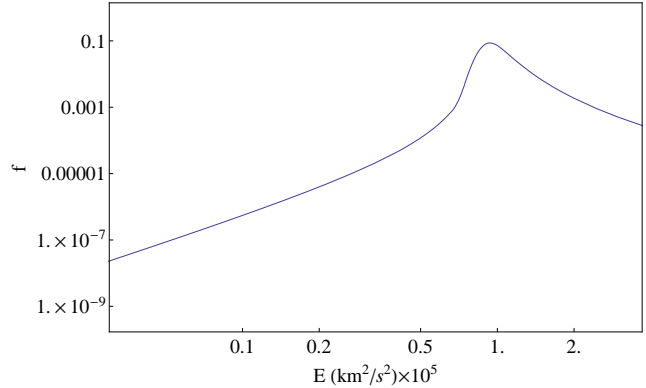}
\caption{Isotropic DF for the two-component spherical model of the NSC
  in the joint gravitational potential including also a central black
  hole. Parameters for the NSC are as given in (\ref{sphParams}), and
  ${M_ \bullet }/(M_1 + M_2)=1.4\times10^{-3}$.}
\label{plot_2}
\end{figure}

The projected number density profile of the NSC obtained from the data
of \cite{fc2014} (see Section~\ref{sDataset}) is shown in Figure \ref{plot_1}. The
inflection point at $R_s \sim 100''$ indicates that the NSC is
embedded in a more extended, lower-density component. The surface
density distribution can be approximated by a two-component model of
the form of equation~(\ref{eqDenSph}), where the six parameters
$({\gamma_1},{M_1},{a_1},\,{\gamma_2},{M_2},{a_2})$ are
fitted to the data subject to the following constraints: The slope of
the inner component should be ${\gamma_1}>0.5$ because isotropic models with a
black hole and ${\gamma_1}<0.5$ are unphysical \citep{tr1994}, but it should be
close to the limiting value of 0.5 to better approximate the observed
core near the center \citep{b2009}. For the outer component
$\gamma_2\ll 0.5$ so that it is negligible in the inner part of the
density profile. In addition $M_1<M_2$ and $a_1< a_2$. With
these constraints we start with some initial values for the
parameters and then iteratively minimize $\chi^2$. The reduced
$\chi^2$ resulting from this procedure is $\chi^2/\nu=0.93$ for $\nu = 55$ d.o.f.
and the corresponding best-fit parameter values are:

\begin{align}
\label{sphParams}
\begin{array}{*{20}{c}}
{{\gamma_1} = 0.51\,}&{a_1 = 99''}\\
{{\gamma_2} = 0.07}&{a_2 = 2376''}
\end{array}\,\,\,\,\frac{{M_2}}{{M_1}} = 105.45.
\end{align}
Here we provide only the ratio of masses instead of absolute values in
model units since the shape of the model depends only on the
ratio. The surface density of the final model is overplotted on the
data in Figure \ref{plot_1}. 

\subsection{Spherical model}

With the assumption of constant mass-to-light
ratio and the addition of the black hole the potential ($\Phi=-\Psi$)
will be \citep{d1993}
\begin{equation}
\begin{array}{l}
{\Psi}(r) = \sum\limits_{i = 1}^2 {\frac{{G{M_i}}}{{{a_i}}}} \frac{1}{{(2 - {\gamma_i})}}
\left( {1 - {{\left( {\frac{r}{{r + a}}} \right)}^{2 - {\gamma_i}}}} \right)+\frac{{G{M_ \bullet }}}{r}
\end{array}
\end{equation}
where ${M_ \bullet }$ is the mass of the black hole. Since we now know
the potential and the density we can calculate the distribution
function (DF) numerically using Eddington's formula, as a function of
positive energy $E=\Psi- \frac{1}{2}{\upsilon^2}$,
\begin{align}
f(E) = \frac{1}{{\sqrt 8 {\pi^2}}}\left[ {{{\int_0^E  {\frac{{d\Psi }}{{\sqrt {E  - \Psi } }}\frac{{{d^2}\rho }}{{d{\Psi^2}}} + \frac{1}{{\sqrt E  }}\left( {\frac{{d\rho }}{{d\Psi }}} \right)} }_{\Psi  = 0}}} \right].
\end{align}
The 2nd term of the equation vanishes for reasonable behavior of the
potential and the double derivative inside the integral can be
calculated easily by using the transformation
\begin{align}
\label{transformation}
\frac{{{d^2}\rho }}{{d{\Psi^2}}} = 
 \left[ { - {{\left( {\frac{{d\Psi }}{{dr}}} \right)}^{ - 3}}\frac{{{d^2}\Psi }}{{d{r^2}}}} \right]
\frac{{d\rho }}{{dr}} + {\left( {\frac{{d\Psi }}{{dr}}} \right)^{ - 2}}\frac{{{d^2}\rho }}{{d{r^2}}}.
\end{align}
Figure \ref{plot_2} shows the DF of the two components in their joint
potential plus that of a black hole with mass ratio ${M_ \bullet
}/(M_1 + M_2)=1.4\times 10^{-3}$. The DF is positive for all
energies. We can test the accuracy of the DF by retrieving the density
using
\begin{align}
\label{eqDFdenSph}
\rho (r) = 4\pi \int\limits_0^\Psi  {dE f(E )\sqrt {\Psi-E}} 
\end{align}
and comparing it with equation~(\ref{eqDenSph}). Both agree to within $0.1\%$.
The DF has the typical shape of models with a shallow cusp of
$\gamma<\frac{3}{2}$. It decreases as a function of energy both in the
neighborhood of the black hole and also for large energies. It has a
maximum near the binding energy of the stellar potential well
\citep{bd2005}.

\begin{figure}
\centering
\includegraphics[width=\linewidth]{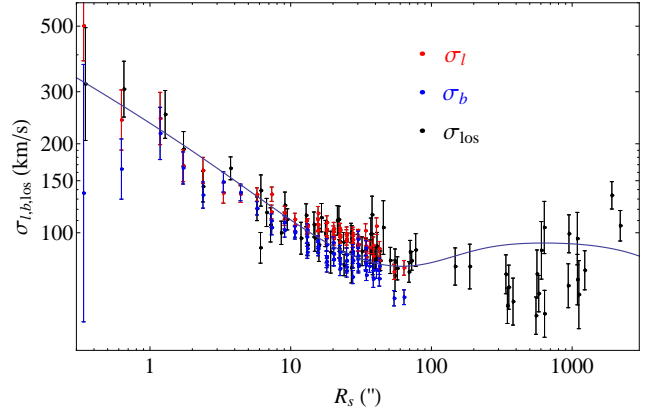}
\caption{Line-of-sight velocity dispersion $\sigma_{\rm los}$ of the two-component
  spherical model with black hole, compared to the observed line-of-sight
  dispersions (black) and the proper motion dispersions in $l$ (red)
  and $b$ (blue). The line-of-sight data includes the outer maser data, and for
  the proper motions a canonical distance of $R_0=8$ kpc is assumed.}
\label{plot_3}
\end{figure}

For a spherical isotropic model the velocity ellipsoid \citep{bt2008}
is a sphere of radius $\sigma$. The intrinsic dispersion $\sigma$ can
be calculated directly using
\begin{align}
\label{eqSDis}
{\sigma^2}(r) = \frac{{4\pi }}{{3\rho (r)}}\int_0^\infty  
{d\upsilon {\upsilon^4}f({\textstyle{1\over2}}{\upsilon^2} - \Psi )}.
\end{align}
The projected dispersion is then given by:
\begin{align}
\label{eqPDis}
\Sigma ({R_s})\sigma_P^2({R_s}) = 2\int_{{R_s}}^\infty  
{{\sigma^2}(r)\frac{{\rho (r)r}}{{\sqrt {{r^2} - R_s^2} }}dr}.
\end{align}
In Figure \ref{plot_3} we see how our two-component model compares
with the kinematical data using the values $R_0 = 8$ kpc for the
distance to the Galactic centre, ${M_\bullet} = 4\times{10^6}{M_ \odot
}$ for the black hole mass, and ${M_*}(r < 100'') =
5\times{10^6}{M_\odot}$ for the cluster mass inside 100''.
The good match of the data up to $80''$ suggests that the
assumption of constant mass-to-light ratio for the cluster is
reasonable. Later-on we will see that a flattened model gives a much
better match also for the maser data.

\subsection{Adding self-consistent rotation to the spherical model}

We describe here the effects of adding self-consistent rotation to the
spherical model, but much of this also applies to the axisymmetric
case which will be discussed in Section~\ref{sAxis}. We assume that the
rotation axis of the NSC is aligned with the rotation axis of the
Milky Way disk. We also use a cartesian coordinate system $(x,y,z)$
where $z$ is parallel to the axis of rotation as before, $y$ is along
the line of sight, and $x$ is along the direction of negative
longitude, with the center of the NSC located at the origin. The
proper motion data are given in Galactic longitude $l$ and Galactic
latitude $b$ angles, but because of the large distance to the center, we
can assume that $x\parallel l$ and $z\parallel b$.

Whether a spherical system can rotate has been answered in
\cite{l1960}. Here we give a brief review. Rotation in a spherical or
axisymmetric system can be added self-consistently by reversing the
sense of rotation of some of its stars. Doing so, the system will
remain in equilibrium. This is equivalent with adding to the DF a part
that is odd with respect to $L_z$. The addition of an odd part does
not affect the density (or the mass) because the integral of the odd
part over velocity space is zero. The most effective way to add
rotation to a spherical system is by reversing the sense of rotation
of all of its counterrotating stars. This corresponds to adding
$f_{-}(E,L^2,L_z) = {\rm sign}({L_z})f(E,L^2)$ \citep[Maxwell's
daemon,][]{l1960} to the initially non-rotating DF, and generates a
system with the maximum allowable rotation. The general case of adding
rotation to a spherical system can be written
$f'(E,L^2,L_z) = (1 + g({L_z}))f(E,L^2)$ where $g({L_z})$ is an odd
function with $\max |g({L_z})| < 1$ to ensure positivity of the DF. We
notice that the new distribution function is a three-integral DF. In
this case the density of the system is still rotationally invariant
but $f_{-}$ is not.

\begin{figure}
\begin{center}
\includegraphics[width=\linewidth]{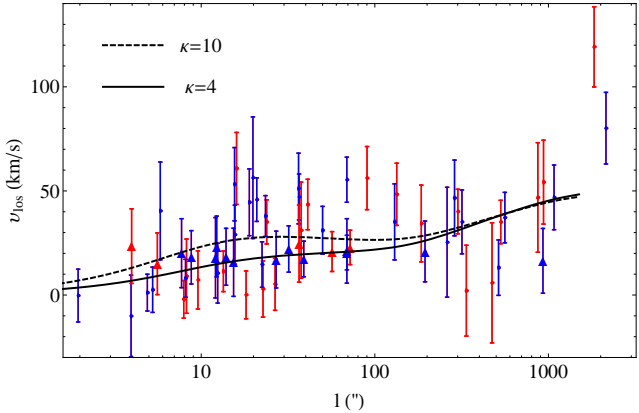}
\end{center}
\caption{Mean line-of-sight velocity data compared to the prediction
of the two-component spherical model with added rotation for
$F=-0.90$ and two $\kappa$ values for illustration. Each data
point corresponds to a cell from Figure \ref{plot_6}.  Velocities at
negative $l$ have been folded over with their signs reversed and are
shown in red. The plot also includes the maser data at $R_s >
100''$.  The model prediction is computed for $b=20''$. For
comparison, cells with centers between $b=15''$ and $b=25''$ are
highlighted with filled triangles.}
\label{plot_4}
\end{figure}

In Figure \ref{plot_3} we notice that the projected velocity
dispersion in the $l$ direction is larger than the dispersion in the
$b$ direction which was first found by \cite{tg2008}. This is
particularly apparent for distances larger than $10''$. A heuristic
attempt to explain this difference was made in \cite{tg2008} where 
they imposed a rotation of the form ${\upsilon_\varphi }(r,\theta)$
along with their Jeans modeling, as a proxy for axisymmetric
modeling. Here we show that for a self-consistent system the
difference in the projected $l$ and $b$ dispersions cannot be
explained by just adding rotation to the cluster.

Specifically, we show that adding an odd part to the distribution
function does not change the proper motion dispersion $\sigma_x$.
The dispersion along the $x$ axis is $\sigma_x^2 = \overline{\upsilon_x^2} -
{\overline\upsilon_x}^2$. Writing $\upsilon_x$ in spherical velocity
components (see the beginning of this section for the notation),
\begin{equation}
\upsilon_x= \upsilon_R {x\over R} - \upsilon_\varphi {y\over R}
 = \upsilon_r\sin\theta {x\over R} + \upsilon_\theta\cos\theta {x\over R} 
    -  \upsilon_\varphi {y\over R}
\end{equation}
we see that
\begin{align}
\begin{array}{l}
\overline{\upsilon_x^2} = \int {d{\upsilon_r}} \int {d{\upsilon_\theta}} 
\int {d{\upsilon_\varphi}}\upsilon_x^2\left( {1 + g({L_z})} \right){f_+ } = \\
 = \int {d{\upsilon_r}} \int {d{\upsilon_\theta}} \int {d{\upsilon_\varphi}} 
       \upsilon_x^2 {f_+ } + 0.
\end{array}
\end{align}
The second term vanishes because $f_+(E,L^2) g(L_z)$ is even in
$\upsilon_r$, $\upsilon_\theta$ and odd in $\upsilon_\varphi$,
so that the integrand for all terms of $f_+ g\, \upsilon_x^2$ is
odd in at least one velocity variable. We also have
\begin{align}
\begin{array}{l}
\overline\upsilon_x= \int {d{\upsilon_r}} \int {d{\upsilon_\theta}} 
\int {d{\upsilon_\varphi}}{\upsilon_x} \left( {1 + g({L_z})} \right){f_ + } = \\
 = 0 - \int {d{\upsilon_r}} \int {d{\upsilon_\theta}} \int {d{\upsilon_\varphi}} 
  {\upsilon_\varphi} {y\over R} {f_+} g.
\end{array}
\end{align}
The first part is zero because ${\upsilon_x}{f_ + }$ is odd. The
second part is different from zero; however when projecting
$\overline\upsilon_\varphi$ along the line-of-sight this term also
vanishes because $f_+ g$ is an even function of $y$ when the
integration is in a direction perpendicular to the $L_z$ angular
momentum direction.  Hence the projected mean velocity
$\overline\upsilon_x$ is zero, and the velocity dispersion
$\sigma_x^2=\overline\upsilon^2_x$ is unchanged.

An alternative way to see this is by making a particle realization of
the initial DF \citep[e.g.][]{ah1974}. Then we can add rotation by
reversing the sign of $L_z$ of a percentage of particles using some
probability function which is equivalent to changing the signs of
$\upsilon_x$ and $\upsilon_y$ of those particles.
$\overline{\upsilon_x^2}$ will not be affected by the sign change and
the $\overline\upsilon_x^2$ averaged over the line-of-sight will be
zero because for each particle at the front of the system rotating in
a specific direction there will be another particle at the rear of the
system rotating in the opposite direction. In this work we do not use
particle models to avoid fluctuations due to the limited number of
particles near the center.

For the odd part of the DF we choose the two-parameter function from
\citet{qh1995}. This is a modified version of \citet{d1986} which was
based on maximum entropy arguments:
\begin{align}
\label{eqAlpha}
g(L_z) = G(\eta ) = F\frac{{\tanh (\kappa \eta /2)}}{{\tanh (\kappa /2)}}
\end{align}
where $\eta = {L_z}/{L_m}(E)$,  ${L_m}(E)$ is the maximum allowable
value of $L_z$ at a given energy, and $-1<F<1$ and $\kappa>0$ are free
parameters. The parameter F works as a global adjustment of rotation
while the parameter $\kappa$ determines the contributions of stars
with different $L_z$ ratios. Specifically for small
$\kappa$ only stars with high $L_z$ will contribute while
large $\kappa$ implies that all stars irrespective of their $L_z$
contribute to rotation. For F=1 and $\kappa \gg 0$,
$g(L_z) = {\rm sign}({L_z})$ which corresponds to maximum rotation.

 From the resulting distribution function $f(E,L_z)$ we can calculate
${\overline\upsilon_\varphi}(R,z)$ in cylindrical coordinates using
the equation
\begin{align}
{\overline\upsilon_\varphi}(R,z) = \frac{{4\pi }}{{\rho {R^2}}}
\int\limits_0^\Psi{dE\int\limits_0^{R\sqrt {2(\Psi - E)} } {{f_ - }(E,{L_z}){L_z}d{L_z}} }.
\label{eqUphi}
\end{align}
To find the mean line-of-sight velocity versus Galactic longitude $l$ we
have to project equation~(\ref{eqUphi}) to the sky plane
\begin{align}
{\upsilon_{\rm los}}(x,z) = \frac{2}{\Sigma }\int_x^\infty  
{{\overline\upsilon_\varphi }(R,z)\frac{x}{R}\frac{{\rho (R,z)RdR}}{{\sqrt {{R^2} - {x^2}} }}}.
\label{ulos}
\end{align}
Figure \ref{plot_4} shows the mean line-of-sight velocity data vs
Galactic longitude $l$ for $F=-0.9$ and two $\kappa$ values for
the parameters in equation~(\ref{eqAlpha}). Later in the axisymmetric section
we constrain these parameters by fitting. Each data point corresponds to
a cell from Figure \ref{plot_6}. The maser data ($r>100''$) are also included.
The signs of velocities for negative $l$ are reversed because of the assumed symmetry.
The line shows the prediction of the model with parameters determined with equation~(\ref{ulos}).
Figure \ref{plot_6} shows that the line-of-sight velocity cells extend from
b=0 to up to $b=50''$, but most of them lie between 0 and
$b=20''$. For this reason we compute the model prediction at an
average value of $b=20''$.

\section[]{AXISYMMETRIC MODELING OF THE NSC}
\label{sAxis}

\begin{figure}
\centering
\includegraphics[width=\linewidth]{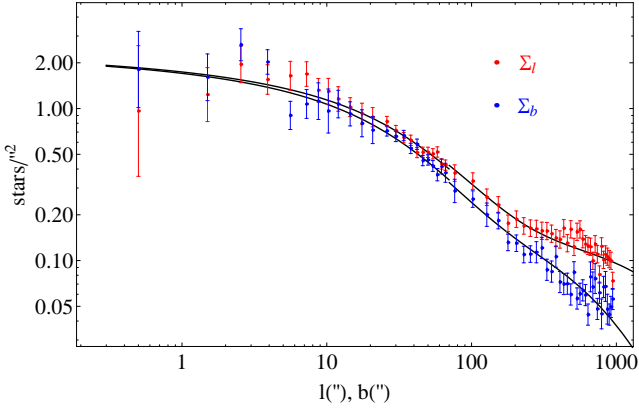}
\caption{Axisymmetric two-component model for the surface density of
  the nuclear cluster. The points with error bars show the number
  density of late-type stars along the $l$ and $b$ directions
  \citep{fc2014} in red and blue respectively. The blue lines show the
  model that gives the best fit to the surface density data with
  parameters as in \ref{bestSDvalues}.}
\label{plot_7}
\end{figure}

We have seen that spherical models cannot explain the difference
between the velocity dispersions along the $l$ and $b$ directions.
The number counts also show that the cluster is flattened; see Figure
\ref{plot_7} and \citet{fc2014}. Therefore we now continue with
axisymmetric modeling of the nuclear cluster. The first step is to fit
the surface density counts with an axisymmetric density model. The
available surface density data extend up to $1000''$ in the $l$ and
$b$ directions. For comparison, the proper motion data extend to $\sim
70''$ from the centre (Figure \ref{plot_5}). We generalize our
spherical two-component $\gamma$-model from equation~(\ref{eqDenSph})
to a spheroidal model given by

\begin{align}
\label{dAxis}
\rho (R,z) = \sum\limits_{i = 1}^2 {\frac{{3 - {\gamma_i}}}
{{4\pi {q_i}}}\frac{{{M_i}\,{a_i}}}{{{m_i^{{\gamma_i}}}{{(m_i + {a_i})}^{4 - {\gamma_i}}}}}} 
\end{align}
where $m_i^2 = {R^2} + {z^2}/q_i^2$ is the spheroidal radius and the
two new parameters $q_{1,2}$ are the axial ratios (prolate $>1$, oblate
$<1$) of the inner and outer component, respectively. Note that the
method can be generalized to N components. The mass of a single
component is given by $4\pi q_i\int\limits_0^\infty {{m_i^2}\rho
  (m_i)dm_i} $. From Figure \ref{plot_7} we expect that the inner
component will be more spherical than the outer component, although
when the density profile gets flatter near the center it becomes more
difficult to determine the axial ratio. In Figure \ref{plot_7} one
also sees that the stellar surface density along the $l$ direction is larger
than along the $b$ direction. Thus we assume that the NSC is an oblate
system. To fit the model we first need to project the density and
express it as a function of $l$ and $b$.  The projected surface density as
seen edge on is
\begin{align}
\label{eqaxisSD}
\Sigma (x,z) = 2\int\limits_x^\infty {\frac{{\rho (R,z)R}}{{\sqrt {{R^2} - {x^2}} }}dR}.
\end{align}

In general, to fit equation~(\ref{eqaxisSD}) to the data we would need
to determine the eight parameters
${\gamma_{1,2}},\,{M_{1,2}},\,{a_{1,2}},\,{q_{1,2}}$. However, we
decided to fix a value for $q_2$ because the second component is not
very well confined in the 8-dimensional parameter space (i.e. there
are several models each with different $q_2$ and similar ${\chi^2}$).
We choose $q_2=0.28$, close to the value found in \cite{fc2014}. For
similar reasons, we also fix the value of ${\gamma_2}$ to that used in
the spherical case. The minimum value of ${\gamma_1}$ for a semi-isotropic
axisymmetric model with a black hole cannot be smaller than $0.5$ \citep{qh1995},
as in the spherical case. For our current modeling we treat ${\gamma_1}$ as a
free parameter. Thus six free parameters remain. To fit these parameters
to the data in Fig.~\ref{plot_7} we apply a Markov chain Monte Carlo algorithm. For comparing the model
surface density (\ref{eqaxisSD}) to the star counts we found it
important to average over angle in the inner conical cells to
prevent an underestimation of the $q_1$ parameter.  The values
obtained with the MCMC algorithm for the NSC parameters and their
errors are:
\begin{align}
\begin{array}{*{20}{c}}
{{\gamma_1} = 0.71 \pm 0.12}&{{a_1} = 147.6'' \pm 27''}&{{q_1} = 0.73 \pm 0.04}\\
{{\gamma_2} = 0.07}&{{a_2} = 4572'' \pm 360''}&{{q_2} = 0.28}\\
  &{{{M_2}}}/{{{M_1}}} = 101.6 \pm 18
\end{array}
\label{bestSDvalues}
\end{align}

The reduced ${\chi^2}$ that corresponds to these parameter values is
${\chi^2}/\nu_{\rm SD} = 0.99$ for $\nu_{\rm SD}=110$ d.o.f.
Here we note that there is a strong correlation between the
parameters $a_2$ and $M_2$.  The flattening of the inner component
is very similar to the recent determination from Spitzer/IRAC
photometry \citep[$0.71\pm0.02$,][]{schodel2014} but slightly more
flattened than the best value given by \citet{fc2014},
$0.80\pm0.04$. The second component is about 100 times more
massive than the first, but also extends more than one order of
magnitude further.

Assuming constant mass-to-light ratio for the star cluster, we
determine its potential using the relation from \cite{qh1995}, which
is compatible with their contour integral method (i.e. it can be used
for complex $R^2$ and $z^2$). The potential for a single component $i$
is given by:
\begin{equation}
\begin{array}{l}
\Psi_i(R,z) = {\Psi_{0i}} - \frac{{2\pi Gq_i}}{e_i}\int\limits_0^\infty  {\rho_i 
\left( U \right)\left[ {\frac{{{R^2}}}{{{{(1 + u)}^2}}} + \frac{{{z^2}}}{{{{({q_i^2} + u)}^2}}}} \right]} \,\\
 \hspace{1.3 cm} \times (\arcsin \,e_i - \arcsin \frac{e_1}{{\sqrt {1 + u} }})du
 \label{ax_pot}
\end{array}
\end{equation}
with $e_i = \sqrt {1 - {q_i^2}}$, $U = \frac{{{R^2}}}{{1 + u}} +
\frac{{{z^2}}}{{{q_i^2} + u}}$, and where $\Psi_{0i}$ is the central
potential (for a review of the potential theory of ellipsoidal bodies
consider \citet{s1969}). The total potential of the two-component model
is
\begin{align}
\Psi (R,z) = \sum\limits_{i = 1}^2 {{\Psi_i}(R,z) + 
\frac{{G{M_ \bullet }}}{{\sqrt {{R^2} + {z^2}} }}}.
\label{ax_tot_pot}
\end{align}

\subsection{Axisymmetric Jeans modeling}

Here we first continue with axisymmetric Jeans modeling. We will need
a large number of models to determine the best values for the mass and
distance of the NSC, and for the mass of the embedded black hole.  We
will use DFs for the detailed modeling in Section 4.3, but this is
computationally expensive, and so a large parameter study with the DF
approach is not currently feasible. In Section~4.3 we will show that a two-integral (2I)
distribution function of the form $f(E,L_z^2)$ gives a very good
representation to the histograms of proper motions and line-of-sight
velocities for the nuclear star cluster in all bins. Therefore we can
assume for our Jeans models that the system is semi-isotropic, i.e., isotropic in the
meridional plane, $\overline{\upsilon_z^2}=\overline{\upsilon_R^2} $.
From the tensor virial theorem \citep{bt2008} we know that for 2I-models $\overline
{\upsilon_\Phi^2}> \overline{\upsilon_R^2}$ in order to produce the
flattening. In principle, for systems of the form $f(E,L_z)$ it is
possible to find recursive expressions for any moment of the
distribution function \citep{m1994} if we know the potential and
the density of the system. However, here we will confine ourselves to
the second moments, since later we will recover the
distribution function. By integrating the Jeans equations we get
relations for the independent dispersions \citep{nm1976}:

\begin{align}
\begin{array}{l}
{\overline {\upsilon_z^2} } (R,z) = {\overline {\upsilon_R^2} }(R,z) =  
- \frac{1}{{\rho (R,z)}}\int_z^\infty  {dz'\rho (R,z')\frac{{\partial \Psi }}{{\partial z'}}} \\
{\overline {\upsilon_\varphi^2}} (R,z) = {\overline {\upsilon_R^2} } (R,z) + 
\frac{R}{{\rho (R,z)}}\frac{{\partial (\rho\overline{\upsilon_R^2} )}} 
{{\partial R}} - R\frac{{\partial \Psi }}{{\partial R}}
\end{array}
\label{nagai}
\end{align}
The potential and density are already known from the previous
section. Once $\overline{\upsilon_z^2}$ is found it can be used to
calculate $\overline{\upsilon_\varphi^2}$. The intrinsic dispersions
in $l$ and $b$ direction are given by the equations:
\begin{align}
\begin{array}{l}
\label{sigma_proj0}
\sigma_b^2 = \overline{\upsilon_z^2}\\
\sigma_l^2 = \overline{ {\upsilon_{x}^2}} = \overline{\upsilon_R^2}{\sin^2}\theta  
             + \overline{\upsilon_\varphi^2}{\cos^2}\theta \\
\overline{ {\upsilon_{\rm los}^2}}  = \overline{ {\upsilon_{y}^2}} 
   = \overline{\upsilon_R^2}{\cos^2}\theta  + \overline{\upsilon_\varphi^2}{\sin^2}\theta 
\end{array}
\end{align}
where ${\sin^2}\theta  = x^2/R^2$ and ${\cos^2}\theta  = 1 - x^2/R^2$.
Projecting the previous equations along the line of sight we have:
\begin{align}
\begin{array}{l}
\label{sigma_proj}
\Sigma \sigma_l^2(x,z) =\\
 2\int_x^\infty  {\left[ {\overline{\upsilon_R^2}\frac{{{x^2}}}{{{R^2}}} + \overline{\upsilon_\varphi^2}\left( {1 - \frac{{{x^2}}}{{{R^2}}}} \right)} \right]
   \frac{{\rho (R,z)}}{{\sqrt {{R^2} - {x^2}} }}dR}, \\
\Sigma \sigma_b^2(x,z) =\\
 2\int_x^\infty  {\overline{\upsilon_z^2}(R,z)\frac{{\rho (R,z)}}{{\sqrt {{R^2} - {x^2}} }}dR}, \\
 \Sigma{\overline{ {\upsilon_{\rm los}^2}}}(x,z)=\\
 2\int_x^\infty {\left[ {\overline{\upsilon_R^2}\left( {1 - \frac{{{x^2}}}{{{R^2}}}} \right) + \overline{\upsilon_\varphi^2} \frac{{{x^2}}}{{{R^2}}}} \right]
   \frac{{\rho (R,z)}}{{\sqrt {{R^2} - {x^2}} }}dR},
\end{array}
\end{align}
where we note that the last quantity in (\ref{sigma_proj0}) and
(\ref{sigma_proj}) is the $2^{nd}$ moment and not the line-of-sight
velocity dispersion.

\begin{figure}
\centering
\includegraphics[width=\linewidth]{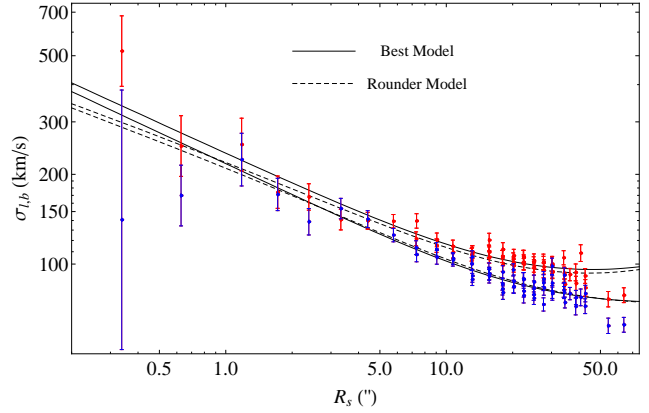}
\caption{ Velocity dispersions $\sigma_l$ and $\sigma_b$ compared to
  axisymmetric, semi-isotropic Jeans models. The measured dispersions
  $\sigma_l$ (red points with error bars) and $\sigma_b$ (blue
  points) for all cells are plotted as a function of their
  two-dimensional radius on the sky, with the Galactic centre at the
  origin.  The black lines show the best model; the model velocity
  dispersions are averaged over azimuth on the sky. The
  dashed black lines show the same quantities for a model which has
  lower flattening ($q_1=0.85$ vs $q_1=0.73$) and a smaller central
  density slope ($0.5$ vs $0.7$).}
\label{plot_9}
\end{figure}

\begin{figure}
\centering
\includegraphics[width=\linewidth]{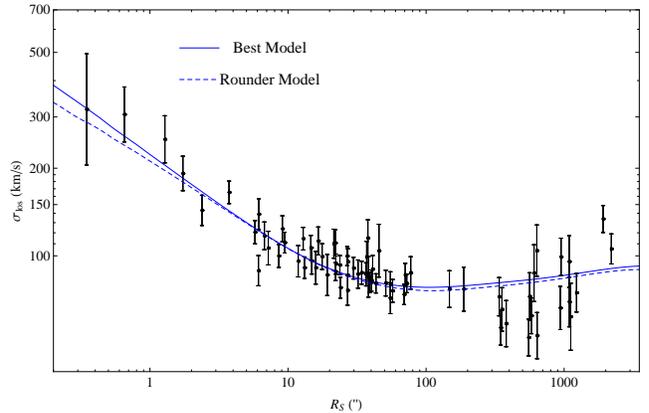}
\caption{Root mean square line-of-sight velocities compared with the
  best model, as a function of two-dimensional radius on
  the sky as in Fig.~\ref{plot_9}. In both plots the stellar mass of
  the NSC is $7.73\times10^6$ ${M_\odot}$ within $m < 100''$, the black
  hole mass is $3.86\times10^6$ ${M_\odot}$, and the distance is $8.3$ kpc
  (equation~\ref{bestModel}). All the maser data are included in the
  plot.}
\label{plot_12}
\end{figure}

In order to define our model completely, we need to determine the distance
$R_0$ and mass $M_*$ of the cluster and the black hole mass
$M_{\bullet}$.  To do this we apply a $\chi^2$ minimization technique
matching all three velocity dispersions in both sets of cells, using
the following procedure. First we note that the inclusion of
self-consistent rotation to the model will not affect its mass. This
means that for the fitting we can use ${\overline{ {\upsilon_{\rm
los}^2} }^{1/2}}$ for each cell of Figure~\ref{plot_6}. 
Similarly, since our model is axisymmetric we should
match to the ${ \overline{\upsilon^2_{l,b}}^{1/2}}$ for each proper
motion cell; the ${{\overline\upsilon_{l,b}}}$ terms should be and
indeed are negligible. Another way to see this is that since the
system is axially symmetric, the integration of $
{{\overline\upsilon_{l,b}}} $ along the line-of-sight should be zero
because the integration would cancel out for positive and negative
$y$.

With this in mind we proceed as follows, using the cluster's density
parameters\footnote{It is computationally too expensive to
  simultaneously also minimize $\chi^2$ over the density parameters.}
as given in (\ref{bestSDvalues}). First we partition the 3d space
($R_0$, $M_*$, $M_{\bullet}$) into a grid with resolution
$20\times20\times20$. Then for each point of the grid we calculate the
corresponding $\chi^2$ using the velocity dispersions from all cells
in Figs.~\ref{plot_5} and \ref{plot_6}, excluding the two cells at the
largest radii (see Fig.~\ref{plot_9}). We compare the measured
dispersions with the model values obtained from
equations~(\ref{sigma_proj}) for the centers of these cells. Then we
interpolate between the $\chi^2$ values on the grid and find the
minimum of the interpolated function, i.e., the best values for ($R_0$,
$M_*$, $M_{\bullet}$). To determine statistical errors on these
quantities, we first calculate the Hessian matrix from the curvature
of $\chi^2$ surface at the minimum, $\partial {\chi^2}/\partial
{p_i}\partial {p_j}$. The statistical variances will be the diagonal
elements of the inverted matrix. 

With this procedure we obtain a minimum reduced $\chi^2/\nu_{\rm
Jeans}=1.07$ with $\nu_{\rm Jeans}=161$ degrees of freedom, for the values
\begin{align}
\begin{array}{l}
{R_0} = 8.27 \, {\rm kpc}\\
{M_*}(m < 100'') = 7.73 \times {10^6}{M_ \odot }\\
{M_ \bullet } = 3.86 \times {10^6}{M_ \odot },
\label{bestModel}
\end{array}
\end{align}
where
\begin{equation}
M_*(m)\equiv \int_0^m 4\pi m^2 \left[ q_1\rho_1(m)+q_2\rho_2(m)\right] dm,
\label{Minsidem}
\end{equation}
and the value given for $M_*$ in (\ref{bestModel}) is not the
total cluster mass but the stellar mass within elliptical radius
$100''$.  In Section~\ref{s_distance} we will consider in more
detail the determination of these parameters and their errors.  The
model with density parameters as in (\ref{bestSDvalues}) and
dynamical parameters as in (\ref{bestModel}) will be our best model.
In Section~\ref{s_2i} we will see that it also gives an excellent
prediction to the velocity histograms.

\begin{figure}
\centering
\includegraphics[width=\linewidth]{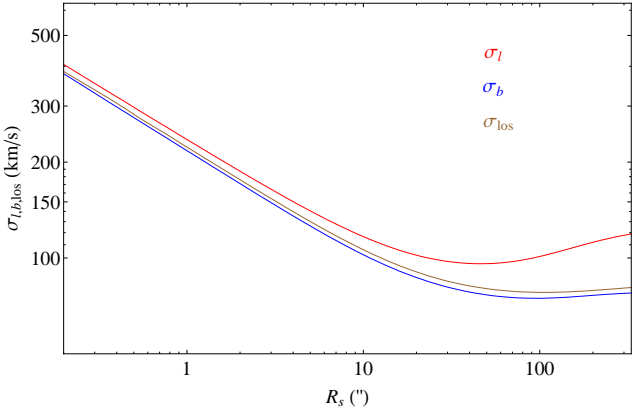}
\caption{All three projected velocity disperions compared. Red:
  ${\sigma_l}$, Blue: ${\sigma_b}$, Brown: ${\sigma
   _{\rm los}}={\overline{ {\upsilon_{\rm los}^2} }^{1/2}}$ .  Note that
  ${\sigma_b}$ is slightly lower than ${\sigma_{\rm los}}$. The
  difference between ${\sigma_b}$ and ${\sigma_l}$ comes from the
  flattening of both the inner and outer components of the model.}
\label{plot_11}
\end{figure}

First, we now look at the comparison of this model with the velocity
data.  Figure~\ref{plot_9} shows how the azimuthally averaged 
dispersions $\sigma_l$ and
$\sigma_b$ compare with the measured proper motion dispersions.
Figure~\ref{plot_12} shows how this best model, similarly
averaged, compares with the line-of-sight mean square velocity
data. The maser data are also included in the plot.  It is seen that
the model fits the data very well, in accordance with its
$\chi^2/\nu_{\rm Jeans}=1.07$ per cell. Figure~\ref{plot_11} shows
how all three projected dispersions of the model compare. $\sigma_{\rm
b}$ is slightly lower than $\sigma_{\rm los}$ due to projection
effects. The fact that all three velocity dispersion profiles in
Figs.~\ref{plot_9},~\ref{plot_12} are fitted well by the model
suggests that the assumed semi-isotropic dynamical structure is
a reasonable approximation.

The model prediction in Fig.~\ref{plot_9} is similar to Figure 11 of
\cite{tg2008} but the interpretation is different.  As shown in the
previous section, the difference in projected dispersions cannot be
explained by imposing rotation on the model. Here we demonstrated how
the observational finding $\sigma_l>\sigma_b$ can be quantitatively
reproduced by flattened axisymmetric models of the NSC and the
surrounding nuclear disk.

Most of our velocity data are in the range 7''-70'', i.e.,
where the inner NSC component dominates the potential. In order to
understand the dynamical implications of these data on the
flattening of this component, we have also constructed several
density models in which we fixed $q_1$ to values different from the
$q_1=0.73$ obtained from star counts. In each case we repeated the
fitting of the dynamical parameters as in (\ref{bestModel}).  We
found that models with $q_1$ in a range from $\sim0.69$ to
$\sim0.74$ gave comparable fits ($\chi^2/\nu$) to the velocity
dispersion data as our nominal best model but that a model with
$q_1=0.77$ was noticeably worse. We present an illustrative model
with flattening about half-way between the measured $q_1=0.73$ and
the spherical case, for which we set $q_1=0.85$. This is also close
to the value given by \citep{fc2014}, $q_1=0.80\pm0.04$.  We
simultaneously explore a slightly different inner slope,
$\gamma_1=0.5$.  We then repeat the fitting of the starcount density
profile in Fig.~\ref{plot_7} (model not shown), keeping also
$\gamma_2$ and $q_2$ fixed to the previous values, and varying the
remaining parameters. Our rounder comparison model then has the
following density parameters:
\begin{align}
\begin{array}{*{20}{c}}
{{\gamma_1} = 0.51}&{{a_1} = 102.6''\,}&{{q_1} = 0.85}\\
{{\gamma_2} = 0.07}&{{a_2} = 4086''}&{{q_2} = 0.28}
\end{array}\,\,\,\frac{{{M_2}}}{{{M_1}}} = 109.1
\label{SDvalues1}
\end{align}
The best reduced $\chi^2$ that we obtain for the velocity dispersion
profiles with these parameters is $\chi^2/\nu_{\rm Jeans}=1.16$ and
corresponds to the values
\begin{align}
\begin{array}{l}
{R_0} = 8.20 \, {\rm kpc}\\
{M_*}(m < 100'') = 8.31 \times {10^6}{M_ \odot }\\
{M_ \bullet } = 3.50 \times {10^6}{M_ \odot },
\label{lessFlattened}
\end{array}
\end{align}

Compared to the best and more flattened model, the cluster mass has
increased and the black hole mass has decreased. The sum of both
masses has changed only by $2\%$ and the distance only by $1\%$.  In
Figures~\ref{plot_9}, \ref{plot_12} we see how the projected velocity
dispersions of this model compare with our best model. The
main difference seen in $\sigma_l$ comes from the different
flattening of the inner component, and the smaller slope of the
dispersions near the center of the new model is because of its
smaller central density slope.

\begin{figure*}
\centering
\includegraphics[width=\linewidth]{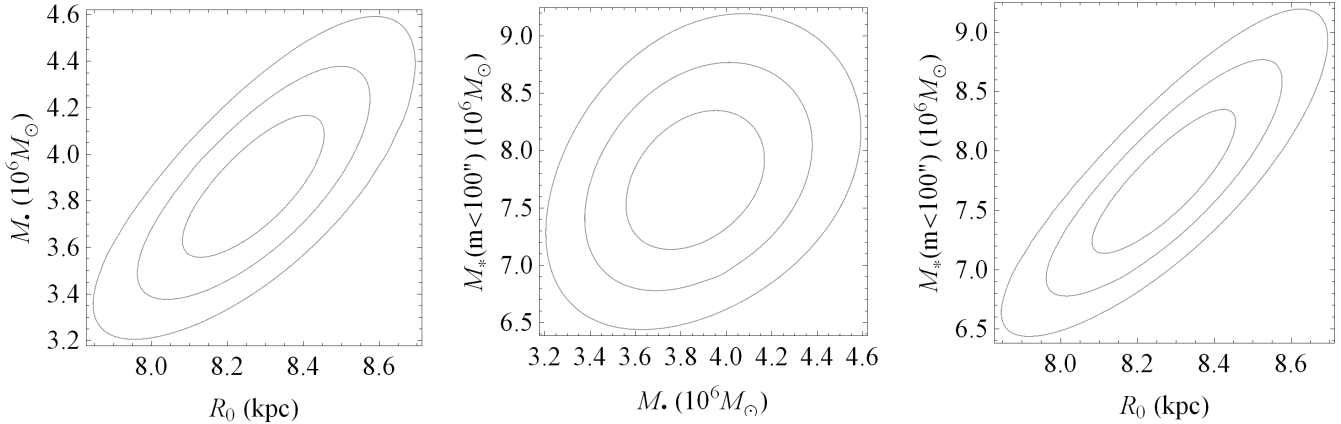}
\caption{Contour plots for the marginalized $\chi^2$ in the three 
  parameter planes $(R_0,M_\bullet)$, $(M_\bullet,M_*)$, $(R_0,M_*)$.
  Contours are plotted at confidence levels corresponding to
  $1\sigma$, $2\sigma$ and $3\sigma$ of the
  joint probability distribution. The minimum corresponds to the values
$R_0 = 8.27 {\rm kpc}$,
$M_*(m < 100'') = 7.73 \times{10^6}{M_\odot}$,
$M_\bullet=3.86 \times {10^6}{M_\odot}$, with errors discussed in
Section~\ref{s_distance}.
}
\label{plot_18}
\end{figure*}

\subsection{Distance to the Galactic Center, mass of the star cluster, and mass of the black hole}
\label{s_distance}

We now consider the determination of these parameters from the NSC
data in more detail.  Fig \ref{plot_18} shows the marginalized
$\chi^2$-plot for the NSC model as given in
equation~(\ref{bestSDvalues}), for pairs of two parameters
$(R_0,M_\bullet)$, $(M_\bullet,M_*)$, $(R_0,M_*)$, as obtained from
fitting the Jeans dynamical model to the velocity dispersion
profiles. The figure shows contour plots for constant $\chi^2/\nu_{\rm
Jeans}$ with $1\sigma$, $2\sigma$ and $3\sigma$ in the three planes for
the two-dimensional distribution of the respective parameters.
We notice that the distance $R_0$ has the smallest relative error.

The best-fitting values for $(R_0,M_*,M_\bullet)$ are given in
equation~(\ref{bestModel}); these values are our best estimates based
on the NSC data alone. For the dynamical model with these
parameters and the surface density parameters given in
(\ref{bestSDvalues}), the flattening of the inner component inferred
from the surface density data is consistent with the dynamical
flattening, which is largely determined by the ratio of
$\sigma_l/\sigma_b$ and the tensor virial theorem.

Statistical errors are determined from the Hessian matrix for this
model.  Systematic errors can arise from uncertainties in the NSC
density structure, from deviations from the assumed axisymmetric
two-integral dynamical structure, from dust extinction within the
cluster (see Section~\ref{s_discussion}), and other sources.  We have
already illustrated the effect of varying the cluster flattening on
$(R_0,M_\bullet,M_*)$ with our second, rounder model.  We have also
tested how variations of the cluster density structure $(a_2,q_2,M_2)$
beyond $500''$ impact the best-fit parameters, and found that these
effects are smaller than those due to flattening variations.

We have additionally estimated the uncertainty introduced by the
symmetrisation of the data if the misalignment found by
\cite{feldmeier2014, fc2014} were intrinsic to the cluster, as
follows.  We took all radial velocity stars and rotated each star by
10$^{\circ}$ clockwise on the sky.  Then we resorted the stars into
our radial velocity grid (Fig.~\ref{plot_6}). Using the new values
${\overline{ {\upsilon_{\rm los}^2} }^{1/2}}$ obtained in the cells
we fitted Jeans models as before. The values we found for $R_0$,
$M_*$, $M_\bullet$ with these tilted data differed from those in
equation~(\ref{bestModel}) by $\Delta R_0=-0.02$ kpc, $\Delta M_*(m
< 100'')=-0.15 \times 10^6 M_\odot$, and $\Delta
M_\bullet=+0.02\times 10^6 M_\odot$, respectively, which are well
within the statistical errors.

Propagating the errors of the surface density parameters from the MCMC
fit and taking into account the correlation of the parameters, we
estimate the systematic uncertainties from the NSC density structure
to be $\sim 0.1$ kpc in $R_0$, $\sim 6\%$ in $M_\bullet$, and $\sim
8\%$ $M_*(m<100'')$.  We will see in Section~\ref{s_2i} below that the
DF for our illustrative rounder NSC model gives a clearly inferior
representation of the velocity histograms than our best kinematic
model, and also that the systematic differences between both models
appear comparable to the residual differences between our preferred
model and the observed histograms.  Therefore we take the differences
between these models, $\sim 10\%$ in ${M_*}$, $\sim10\%$ in
${M_\bullet }$, and $\sim0.1 {\rm kpc}$ in $R_0$, as a more
conservative estimate of the dynamical modeling uncertainties, so that
finally

\begin{align}
\begin{array}{l}
{R_0} = 8.27 \pm 0.09{|_{\rm stat}}\pm 0.1{|_{\rm syst}} \, {\rm kpc}\\
{M_*}(m < 100'') = (7.73 \pm 0.31{|_{\rm stat}}\pm 0.8{|_{\rm syst}}) \times {10^6}{M_\odot }\\
{M_ \bullet } = (3.86 \pm 0.14{|_{\rm stat}\pm 0.4{|_{\rm syst}}}) \times {10^6}{M_\odot }.
\label{bestflattenedWithErrors}
\end{array}
\end{align}

We note several other systematic errors which are not easily
quantifiable and so are not included in these estimates, such as
inhomogeneous sampling of proper motions or line-of-sight velocities,
extinction within the NSC, and the presence of an additional component
of dark stellar remnants.

Based on our best model, the mass of the star cluster within $100''$
converted into spherical coordinates is ${M_*}(r < 100'') = (8.94 \pm
0.32{|_{\rm stat}} \pm 0.9{|_{\rm syst}}) \times {10^6}{M_\odot }$.
The model's mass within the innermost pc ($25''$) is ${M_*}(m < 1{\rm
  pc}) = 0.729\times{10^6}{M_\odot }$ in spheroidal radius, or
${M_*}(r < 1{\rm pc}) = 0.89\times{10^6}{M_\odot }$ in spherical
radius.  The total mass of the inner NSC component is ${M_{1}} = 6.1
\times {10^7}{M_\odot }$.  Because most of this mass is located beyond
the radius where the inner component dominates the projected star
counts, the precise division of the mass in the model between
the NSC and the adjacent nuclear disk is dependent on the assumed
slope of the outer density profile of NSC, and is therefore
uncertain.

\begin{figure}
\centering
\includegraphics[width=\linewidth]{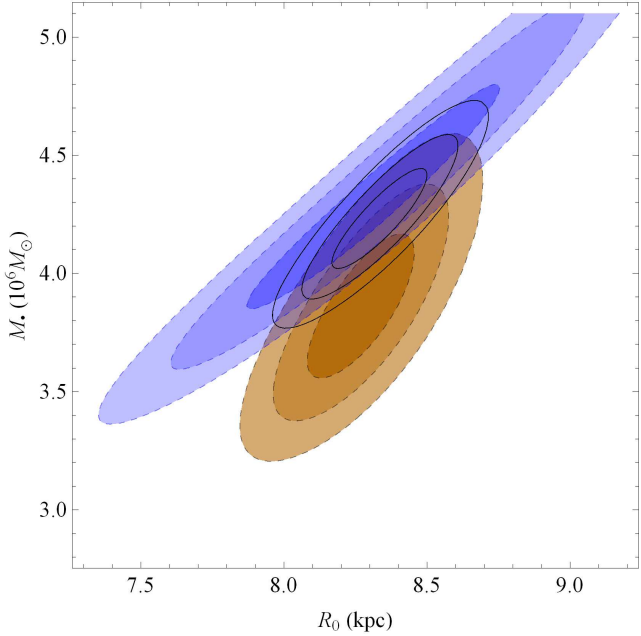}
\caption{Blue: $\chi^2$ contours in the $(R_0, M_{\bullet})$ plane
  from stellar orbits of S-stars, as in Figure 15 of \citet{ge2009},
  at confidence levels corresponding to $1\sigma$, $2\sigma$,
  $3\sigma$ for the joint probability distribution.  Brown:
  Corresponding $\chi^2$ contours from this work. Black: Combined
  contours after adding the $\chi^2$ values.}
\label{plot_22}
\end{figure}

The distance and the black hole mass we found differ by $0.7\%$ and
$12\%$, respectively, from the values $R_0 = 8.33 \pm 0.17{|_{\rm
stat}} \pm 0.31{|_{\rm syst}}$ kpc and ${M_{\bullet}} = 4.31 \pm
0.36\times{10^6}{M_\odot }$ for $R_0=8.33$ kpc, as determined by
\cite{ge2009} from stellar orbits around Sgr A$^*$.
Figure~\ref{plot_22} shows the $1\sigma$ to $3\sigma$ contours of
marginalized $\chi^2$ for $(R_0, M_\bullet)$ jointly from stellar
orbits \citep{ge2009}, for the NSC model of this paper, and for the
combined modeling of both data sets. The figure shows that both
analyses are mutually consistent. When marginalized over $M_*$ and the respective other parameter, 
the combined modeling gives, for each parameter alone, 
$R_0=8.33\pm0.11$ kpc and $M_\bullet=4.23\pm0.14 \times 10^6 M_\odot$.
We note that these errors for $R_0$ and $M_\bullet$ are both dominated by the
distance error from the NSC modeling. Thus our estimated additional
systematic error of $0.1$ kpc for $R_0$ in the NSC modeling translates to a
similar additional error in the combined $R_0$ measurement and, through
the SMBH mass-distance relation given in Gillessen et al (2009), to an
additional uncertainty $\simeq0.1\times10^6 M_\odot$ in $M_\bullet$.
We see that the combination of the NSC and S-star orbit data is a
powerful means for decreasing the degeneracy between the SMBH mass and
Galactic center distance in the S-star analysis.

\begin{figure*}
\centering
\includegraphics[width=\linewidth]{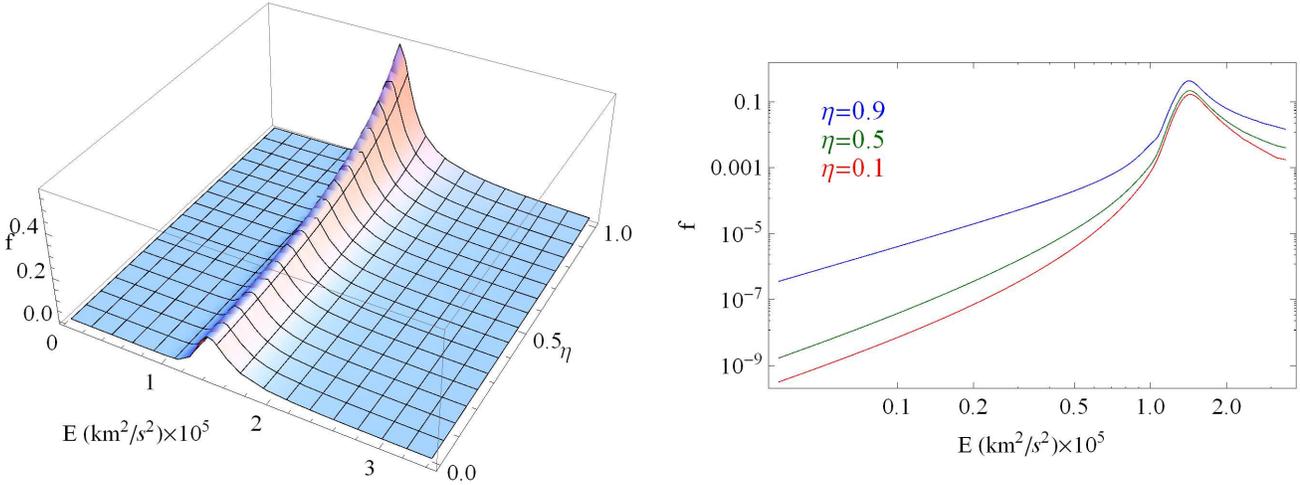}
\caption{We used the HQ algorithm to calculate the 2I-DF for our best
  Jeans model. The left plot shows the DF in $E$ and $\eta =
  {L_z}/{L_{z\max }}(E)$ space. The DF is an increasing function of
  $\eta$. The right plot shows the projection of the DF on energy
  space for several values of $\eta$. The shape resembles that of the
  spherical case in Figure \ref{plot_2}. }
\label{plot_10}
\end{figure*}

\subsection[]{Two-integral distribution function for the NSC.}
\label{s_2i}

Now we have seen the success of fitting the semi-isotropic Jeans
models to all three velocity dispersion profiles of the NSC, and
determined its mass and distance parameters, we proceed to calculate
two-integral (2I) distribution functions.  We use the contour integral
method of \citet[][HQ]{hq1993} and \citet{qh1995}.  A 2I DF is the
logical, next-simplest generalization of isotropic spherical
models. Finding a positive DF will ensure that our model is
physical. Other possible methods to determine $f(E,L_z$) include
reconstructing the DF from moments \citep{m1995}, using series
expansions as in \cite{dg1994}, or grid-based quadratic programming as
in \cite{k1995}.  We find the HQ method the most suitable since it is
a straightforward generalization of Eddington's formula. The contour
integral is given by:
\begin{align}
\begin{array}{l}
{f_+}(E,{L_z}) =\\
\frac{1}{{4{\pi^2}i\sqrt 2 }}\oint 
{\frac{{d\xi }}{{{{(\xi  - E)}^{1/2}}}}{\tilde\rho_{11}}\left( {\xi ,\frac{{L_z^2}}{{2{{(\xi  - E)}^{1/2}}}}} \right)}
\label{HQDF}
\end{array}
\end{align}
where ${\tilde\rho_{11}}(\Psi ,R) = \frac{{{\partial^2}}}{{\partial
    {\Psi^2}}}\rho (\Psi ,R)$. Equation~(\ref{HQDF}) is remarkably
similar to Eddington's formula. Like in the spherical case the DF is
even in $L_z$. The integration for each $(E,{L_z})$ pair takes place
on the complex plane of the potential $\xi$ following a closed path
(i.e. an ellipse) around the special value ${\Psi_{\rm env}}$. For
more information on the implementation and for a minor improvement
over the original method see Appendix A.  We find that a resolution of
$(120\times60)$ logarithmically placed cells in the $(E,{L_z})$ space
is adequate to give us relative errors of the order of $10^{-3}$ when
comparing with the zeroth moment, i.e., the density, already known
analytically, and with the second moments, i.e., the velocity
dispersions from Jeans modeling.

The gravitational potential is already known from
equations~(\ref{ax_pot}) and (\ref{ax_tot_pot}). For the parameters
(cluster mass, black hole mass, distance) we use the values given in
(\ref{bestModel}).  Figure \ref{plot_10} shows the DF in
$(E,{L_z})$ space. The shape resembles that of the spherical case
(Fig.~\ref{plot_2}). The DF is a monotonically increasing function of
$\eta = {L_z}/{L_{z\max }}(E)$ and declines for small and large
energies. The DF contains information about all moments and therefore
we can calculate the projected velocity profiles (i.e., velocity
distributions, hereafter abbreviated VPs) in all directions.
The normalized VP in the line-of-sight (los) direction $y$ is
\begin{align}
VP({\upsilon_{\rm los}};x,z) = \frac{1}{\Sigma }\iiint\limits_{E>0} {f(E,{L_z})\,d{\upsilon_{x}}d{\upsilon_z}dy}.
\label{VPlos}
\end{align}

Using polar coordinates in the velocity space
$({\upsilon_x},{\upsilon_z}) \to ({\upsilon_ \bot },\varphi )$ where
${\upsilon_x} = {\upsilon_ \bot }\cos \varphi$ and ${\upsilon_z} =
{\upsilon_ \bot }\sin \varphi$ we find

\begin{align}
VP({\upsilon_{\rm los}};x,z) = \frac{1}{{2\Sigma }}\int\limits_{{y_1}}^{{y_2}} {dy} 
\int\limits_0^{2\Psi-\upsilon_{\rm los}^2} {d\upsilon_ \bot^2} \int\limits_0^{2\pi } {\,d\varphi } f(E ,{L_z})
\label{VPlos1}
\end{align}
where
\begin{align}
\begin{array}{l}
E=\Psi (x,y,z) - \frac{1}{2}(\upsilon_{\rm los}^2 + \upsilon_\bot^2),\\
{L_z} = x{\upsilon_{\rm los}} - y{\upsilon_ \bot }\cos \varphi.
\end{array}
\end{align}
and $y_{1,2}$ are the solutions of $\Psi(x,y,z)-v^2_{\rm los}/2=0$.
Following a similar path we can easily find the corresponding
integrals for the VPs in the $l$ and $b$ directions.

\begin{figure}
\centering
\includegraphics[width=\linewidth]{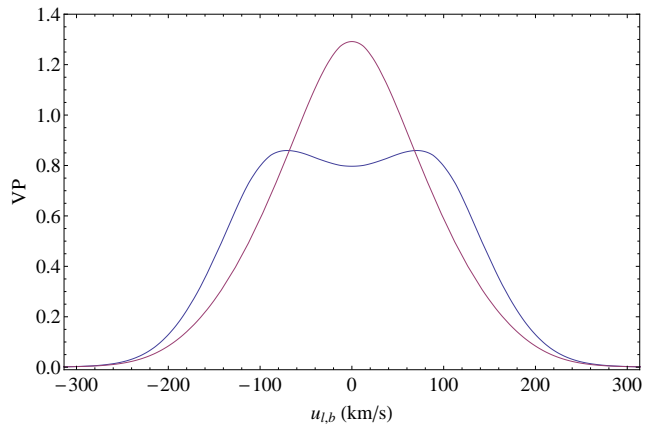}
\caption{Typical velocity distributions for $l$ and $b$-velocities
  within the area of interest $(r<100'')$. The red line shows the VPs
  in the $b$ direction, the blue line in the $l$ direction. The VPs
  along $l$ show the characteristic two-peak-shape pointed out from the
  data by several authors \citep{se2007,tg2008,fc2014}.}
\label{plot_14}
\end{figure}

The typical shape of the VPs in the $l$ and $b$ directions within the
area of interest $(r<100'')$ is shown in Figure \ref{plot_14}. We
notice the characteristic two-peak shape of the VP along $l$ that is
caused by the near-circular orbits of the flattened system.  Because
the front and the back of the axisymmetric cluster contribute equally,
the two peaks are mirror-symmetric, and adding rotation would not
change their shapes.

The middle panels of Figure~\ref{plot_23} and Figures~\ref{plot_15}
and \ref{plot_16} in Appendix B show how our best model (with
parameters as given in~(\ref{bestSDvalues}) and (\ref{SDvalues1}))
predicts the observed velocity histograms for various combinations of
cells. The reduced $\chi^2$ for each histogram is also provided.  The
prediction is very good both for the VPs in $\upsilon_l$ and
$\upsilon_b$. Specifically, for the $l$ proper motions our flattened
cluster model predicts the two-peak structure of the data pointed out by
several authors \citep{tg2008,sm2009,fc2014}. In
order to calculate the VP from the model for each cell we averaged
over the VP functions for the center of each cell weighted by the number
of stars in each cell and normalized by the total number of stars in
all the combined cells.

Figure \ref{plot_23} compares two selected $\upsilon_l$-VPs for our two
main models with the data. The left column shows how the observed velocity
histograms (VHs) for corresponding cells compare
to the model VPs for the less flattened model with parameters given in (\ref{SDvalues1}) and (\ref{lessFlattened}),
the middle column compares with the same VPs from our best model with parameters given
in (\ref{bestSDvalues}) and (\ref{bestModel}). Clearly, the
more flattened model with $q_1=0.73$ fits the shape of the data much better than
the more spherical model with $q_1=0.85$, justifying its use in
Section~\ref{s_distance}.

This model is based on an even DF in $L_z$ and therefore does not yet
have rotation. To include rotation, we will (in Section 4.4) add an
odd part to the DF, but this will not change the even parts of the
model's VPs.  Therefore, we can already see whether the model is also
a good match to the observed los velocities by comparing it to the
even parts of the observed los VHs. This greatly simplifies the
problem since we can think of rotation as independent, and can
therefore adjust it to the data as a final step. Figure \ref{plot_17}
shows how the even parts of the VHs from the los data compare with the
VPs of the 2I model.  Based on the reduced $\chi^2$, the model
provides a very good match.  Possible systematic deviations are within
the errors. The los VHs are broader than those in the $l$ direction
because the los data contain information about rotation (the broader
the even part of the symmetrized los VHs, the more rotation the system
possesses, and in extreme cases they would show two peaks).

\begin{figure*}
\centering
\includegraphics[width=\linewidth]{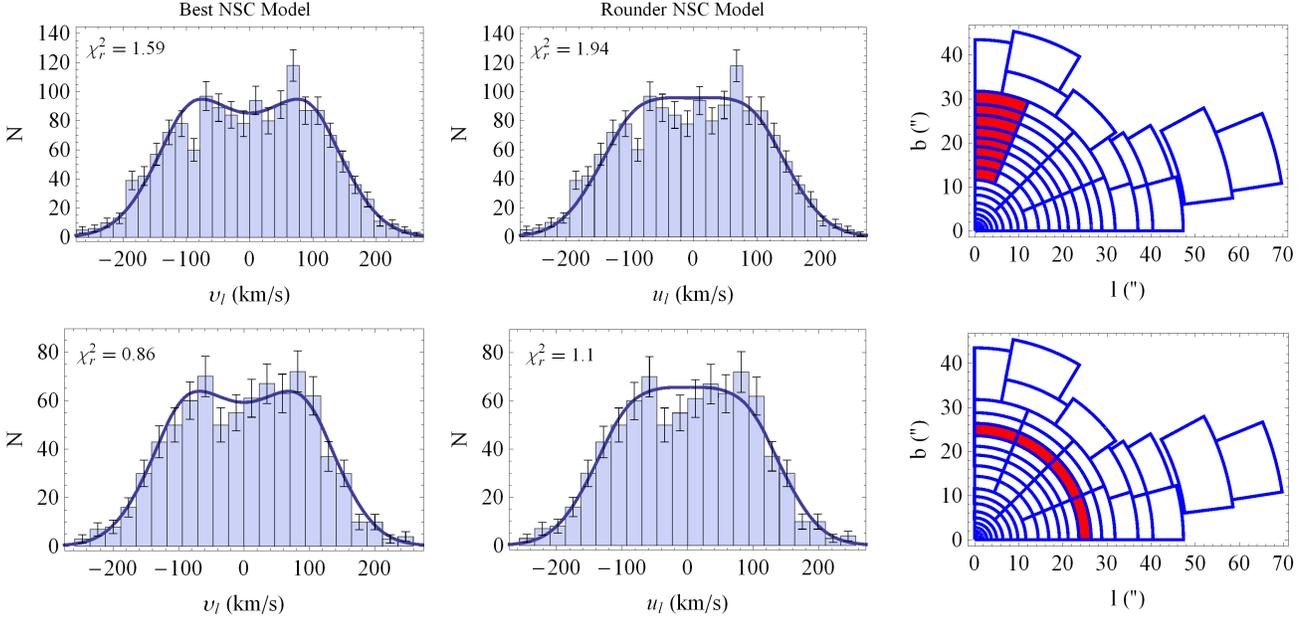}
\caption{Predicted distributions of $\upsilon_l$ velocity compared to the
observed histograms. In each row, model VPs and observed VHs are shown
averaged over the cells indicated in red in the right column,
respectively.  Left column: predictions for the less flattened model
which we use as an illustration model, i.e., for parameters given in
(26) and (27).  Middle column: predicted VPs for our best
model with parameters given in (18) and
(24). This more flattened model with $q_1 = 0.73$ fits the data
much better than the rounder cluster model with $q_1 = 0.85$.}
\label{plot_23}
\end{figure*}

\subsection{Adding rotation to the axisymmetric model: is the NSC an isotropic rotator?}

As in the spherical case, to model the rotation we add an odd part in
$L_z$ to the initial even part of the distribution function, so that
the final DF takes the form $f(E,{L_z}) = (1 + g(L_z) )f(E,{L_z})$.
We use again equation~(\ref{eqAlpha}); this adds two additional
parameters ($\kappa$, F) to the DF. Equation~(\ref{ulos}) gives the
mean los velocity vs Galactic longitude. In order to constrain the
parameters ($\kappa$, F) we fitted the mean los velocity from
equation~(\ref{ulos}) to the los velocity data for all cells in
Fig.~\ref{plot_6}.  The best parameter values resulting from this
2D-fitting are $\kappa=2.8\pm1.7$, $F=0.85\pm0.15$ and
$\chi_{r}^2=1.25$. Figure \ref{plot_19} shows that the VPs of this
rotating model compare well with the observed los VHs.

\begin{figure}
\centering
\includegraphics[width=\linewidth]{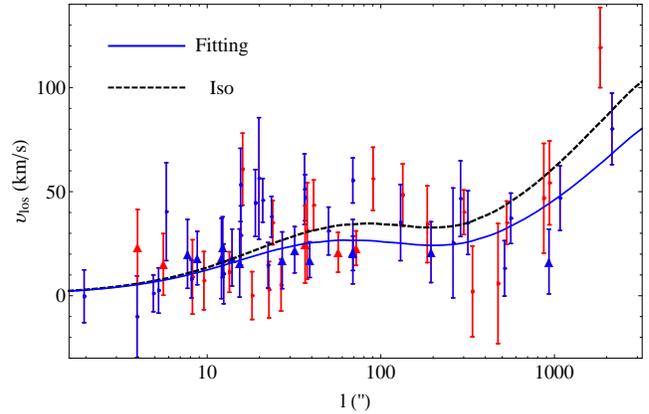}
\caption{Best fitting model from the 2I DF compared to the isotropic
  rotator model.  Each data point corresponds to a cell from Figure
  \ref{plot_6}. Velocities at negative $l$ have been folded over with
  their signs reversed and are shown in red. The plot also includes
  the maser data at $R_s > 100''$. The predictions of both models are
  computed for $b=20''$. For comparison, cells with centers between
  $b=15''$ and $b=25''$ are highlighted with full triangles. }
\label{plot_13}
\end{figure}

An axisymmetric system with a DF of the form $f(E,{L_z})$ is an
isotropic rotator when all three eigenvalues of the dispersion tensor
are equal \citep{bt2008} and therefore
\begin{align}
\overline \upsilon_\varphi^2 =\overline{\upsilon_\varphi^2}-\overline{\upsilon_R^2}.
\label{isotropic}
\end{align}
In order to calculate $\overline \upsilon_\varphi$ from
equation~(\ref{isotropic}) it is not necessary for the DF to be known
since $\overline{\upsilon_\varphi^2}$ and $\overline{\upsilon_R^2}$
are already known from the Jeans equations~(\ref{nagai}). Figure
\ref{plot_13} shows the fitted $u_{\rm los}$ velocity from the DF
against the isotropic rotator case calculated from
equation~(\ref{isotropic}), together with the mean los velocity data.
The two curves agree well within $\sim30''$, and also out to $\sim
200''$ they differ only by $\sim 10$ km/s. Therefore according to our best
model the NSC is close to an isotropic rotator, with slightly
lower rotation and some tangential anisotropy outwards of 30''.

\section{Discussion}
\label{s_discussion}

In this work we presented a dynamical analysis of the Milky Way's
nuclear star cluster (NSC), based on $\sim$10'000 proper motions,
$\sim$2'700 radial velocities, and new star counts from the companion
paper of \cite{fc2014}. We showed that an excellent representation of
the kinematic data can be obtained by assuming a constant
mass-to-light ratio for the cluster, and modeling its dynamics with
axisymmetric two-integral distribution functions (2I-DFs), $f(E,L_z)$.
The DF modeling allows us to see whether the model is physical, i.e.,
whether the DF is positive, and to model the proper motion (PM) and
line-of-sight (los) velocity histograms (VHs). One open question until now has been
the nature of the double peaked VHs of the $v_l$-velocities along
Galactic longitude, and the bell-shaped VHs of $v_b$ along Galactic
latitude, which cannot be fitted by Gaussians \citep{sm2009}.  Our 2I
DF approximation of the NSC gives an excellent prediction for the
observed shapes of the $v_l$-, $v_b$, and $v_{\rm los}$-VHs. The models
show that the double-peaked shape of the $v_l$-VHs is a result of the
flattening of the NSC, and suggest that the cluster's dynamical
structure is close to an isotropic rotator. Because both PMs and los-velocities
enter the dynamical models, we can use them also to constrain the distance to
the GC, the mass of the NSC, and the mass of the Galactic centre black hole. To do this 
efficiently, we used the semi-isotropic Jeans equations corresponding
to 2I-DFs. In this section, we discuss these issues in more detail.

\subsection{The dynamical structure of the NSC}

The star count map derived in \cite{fc2014} suggests two components in
the NSC density profile, separated by an inflection point at about
$\sim200''\sim 8$ pc (see Fig.~\ref{plot_7} above). To account for
this we constructed a two-component dynamical model for the star
counts in which the two components are described as independent
$\gamma$-models. The inner, rounder component can be considered as the
proper NSC, as in \cite{fc2014}, while the outer, much more flattened
component may represent the inner parts of the nuclear stellar disk
(NSD) described in \citet{laun2002}.

The scale radius of the inner component is $\sim 100''$, close to the
radius of influence of the SMBH, $r_h\sim 90''$ \citep{a2005}. The
profile flattens inside $\sim 20''$ to a possible core
\citep{b2009,fc2014} but the slope of the three-dimensional density
profile for the inner component is not well-constrained. 

The flattening for the inner NSC component inferred from star counts
is $q_1=0.73 \pm 0.04$, very close to the value of $q=0.71
\pm 0.02 $ found recently from Spitzer multi-band photometry
\citep{schodel2014}.  It is important that these determinations
agree with the dynamical flattening of our best Jeans dynamical
models: the dynamical flattening is robust because it is largely
determined by the ratio of $\sigma_b/\sigma_l$ and the tensor virial
theorem. Because star counts, photometric, and dynamical values for
the inner NSC flattening agree, this parameter can now be considered
securely determined.

\begin{figure}
\centering
\includegraphics[width=\linewidth]{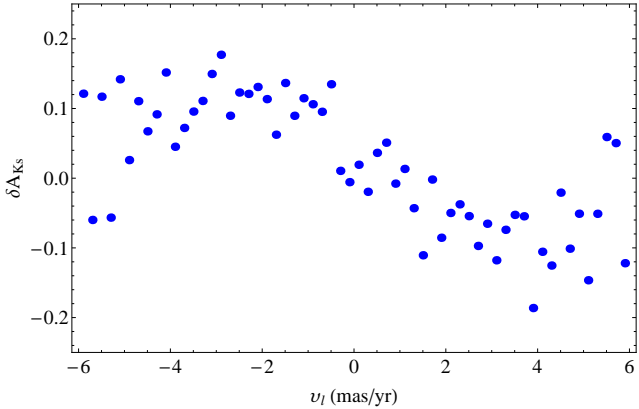}
\caption{Average differential extinction of nuclear cluster stars
plotted as a function of $v_l$ proper motion.  The differential
extinction is inferred from the difference in the color of a star to
the median color of its 16 nearest neighbours, using the extinction
law of \citet{fc2011}, and correcting also for the weak color
variation with magnitude. For this plot we use all the
proper motion stars in the central and extended fields of
\citet{fc2014} and bins of 0.2 mas/yr. The differential
extinction is larger for stars with negative $l$-proper motions
which occur preferentially at the back of the cluster.}
\label{plot_dust}
\end{figure}

Assuming constant mass-to-light ratio for the NSC, we found that a
2I-DF model gives an excellent description of the proper motion and
los velocity dispersions and VHs, in particular of the double-peaked
distributions in the $v_l$-velocities. This double-peaked structure is
a direct consequence of the flattening of the star cluster; the
detailed agreement of the model VPs with the observed histograms
therefore confirms the value $q_1=0.73$ for the inner cluster
component. For an axisymmetric model rotation cannot be seen
directly in the proper motion VHs when observed edge-on, as is the
case here, but is apparent only in the los velocities. When a
suitable odd part of the DF is added to include rotation, the 2I-DF
model also gives a very good representation of the skewed los VHs.
From the amplitude of the required rotation we showed that the NSC can
be approximately described as an isotropic rotator model, 
rotating slightly slower than that outside $\sim30''$.

Individual VHs are generally fitted by this model within the
statistical errors, but on closer examination the combined $v_l$ VHs
show a slightly lower peak at negative velocities, as already apparent
in the global histograms of \citet{tg2008, sm2009}.
Fig.~\ref{plot_dust} suggests that differential extinction of order
$\sim 0.2$ mag within the cluster may be responsible for this small
systematic effect, by causing some stars from the back of the cluster
to fall out of the sample.  The dependence of mean extinction on $v_l$
independently shows that the NSC must be rotating, which could
otherwise only be inferred from the los velocities. In subsequent
work, we will model the effect of extinction on the inferred dynamics
of the NSC. This will then also allow us to estimate better how
important deviations from the 2I-dynamical structure are, i.e.,
whether three-integral dynamical modeling \citep[e.g.,][]{dl2013}
would be worthwhile.

\subsection{Mass of the NSC}
\label{massDiscussion}
\begin{figure}
\centering
\includegraphics[width=\linewidth]{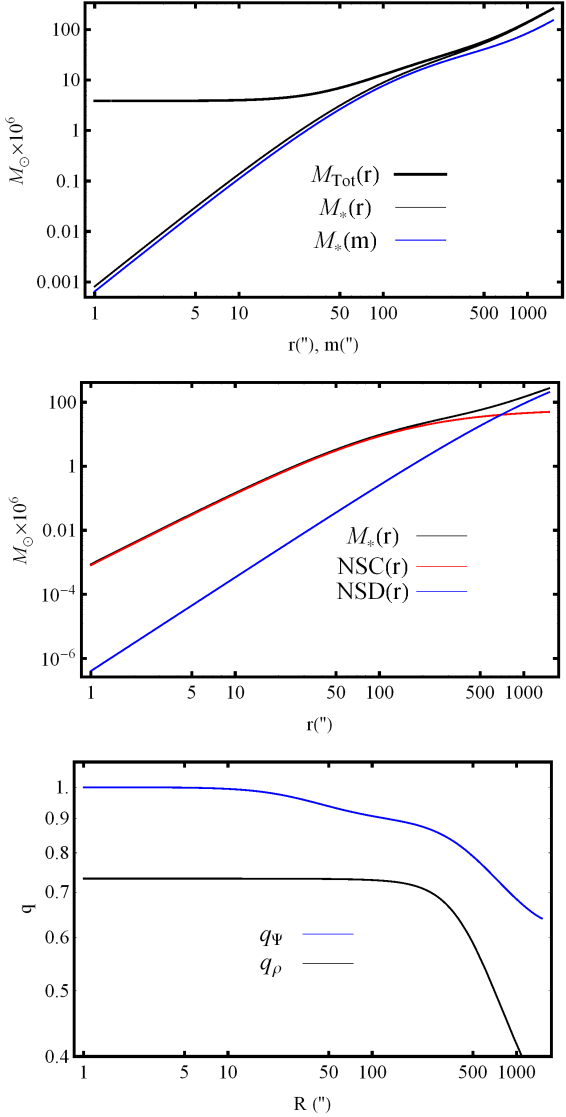}
\caption{Upper panel: Enclosed mass of the NSC, as function of
  three-dimensional radius $r$ and spheroidal radius $m$, and total
  enclosed mass including the black hole.  Middle panel: Enclosed mass
  of the inner component of the NSC (inner component $M_1$), the NSD
  (outer component $M_2$), and total enclosed stellar mass, as
  function of three-dimensional radius $r$. Lower panel: Axis ratios
  of the stellar density and total potential as functions of the
  cylindrical radius $R$. }
\label{plot_8}
\end{figure}

The dynamical model results in an estimate of the mass of the cluster
from our dataset. Our fiducial mass value is ${M_*}(m\!<\! 100'') =
(7.73 \pm 0.31{|_{\rm stat}}\pm 0.8{|_{\rm syst}}) \times
{10^6}{M_\odot }$ interior to a spheroidal major axis distance
$m\!=\!100''$. This corresponds to an enclosed mass within
3-dimensional radius $r\!=\!100''$ of ${M_*}(r\!<\!100'')=(8.94\pm
0.31{|_{\rm stat}}\pm0.9{|_{\rm syst}}) \times {10^6}{M_ \odot}$. 

The fiducial mass ${M_*}(r\!<\!100'')$ for the best axisymmetric model
is larger than that obtained with spherical models. The constant M/L
spherical model with density parameters as in Section 3, for 
$R_0=8.3$ kpc and the same black hole mass has ${M_*}(r\!<\!100'')=6.6
\times 10^6M_\odot$. 

There are two reasons for this difference: (i) At $\sim 50''$ where the model
is well-fixed by kinematic data the black hole still contributes more
than half of the interior mass.  In this region, flattening the
cluster at constant mass leaves $\sigma_l$ and $\sigma_{\rm los}$
approximately constant, but decreases $\sigma_b$ to adjust to the shape. To
fit the same observed data, the NSC mass must be increased. (ii)
Because of the increasing flattening with radius, the average density
of the axisymmetric model decreases faster than that of the spherical
density fit; thus for the same observed velocity dispersion
profiles a larger binding mass for the NSC is required.

Figure~\ref{plot_8} shows the enclosed stellar mass within the spheroidal
radius $m$ as in equation~(\ref{Minsidem}), as well as the mass within
the spherical radius $r$. E.g., the mass within $1$~pc ($25''$) is
${M_*}(r\!<\!1{\rm pc \sim 25''})=0.89 \times {10^6}{M_\odot}$. This
is compatible with the spherical modeling of \cite{sm2009} who gave a
range of $0.6-1.7\times 10^6M_\odot$, rescaled to $R_0=8.3$ kpc, with
the highest mass obtained for their isotropic, constant M/L model.
According to Fig.~\ref{plot_8}, at $r\simeq 30''=1.2$ pc the NSC
contributes already $\simeq 25\%$ of the interior mass ($\simeq 45\%$
at $r\simeq 50''=2$ pc), and beyond $r\simeq 100''=4$ pc it clearly
dominates.

An important point to note is that the cluster mass does not depend on
the net rotation of the cluster but only on its flattening. This is
because to add rotation self-consistently to the model we need to add
an odd part to the DF which does not affect the density or the proper
motion dispersions $\sigma_l$ and $\sigma_b$.

Our NSC mass model can be described as a superposition of a moderately
flattened nuclear cluster embedded in a highly flattened nuclear disk.
The cumulative mass distributions of the two components are shown in
the middle panel of Figure~\ref{plot_8}. The NSD starts to
dominate at about $800''$ which is in good agreement with the value
found by \citet{laun2002}.

Approximate local axis ratios for the combined density and for the
total potential including the central black hole are shown in the
lower panel of Fig.~\ref{plot_8}. Here we approximate the axial
ratio of the density at radius $R$ by solving the equation
$\rho(R,0)=\rho(0,z)$ for z and writing $q_{\rho}=z/R$, and
similarly for $q_\Psi$. The density axis ratio ${q_\rho}{(R)}$ shows
a strong decrease between the regions dominated by the inner and
outer model components. The equipotentials are everywhere less
flattened.  At the center, $q_\Psi = 1$ because of the black hole;
the minimum value is not yet reached at $1000''$.  Therefore, we
can define the NSC proper as the inner component of this model,
similar to \cite{fc2014}.

The total mass of the inner component,
$M_{1}=6.1\times{10^7}{M_\odot}$ (Section~\ref{s_distance}), is
well-determined within similar relative errors as $M_\ast(m\!<\!
100'')$. However, identifying $M_{1}$ with the total mass of the
Galactic NSC at the center of the nuclear disk has considerable
uncertainties: because the outer NSD component dominates the
surface density outside $100''-200''$, the NSC density profile slope
at large radii is uncertain, and therefore the part of the mass
outside $\sim200''$ ($\sim 64\%$ of the total) is also uncertain.  A
minimal estimate for the mass of the inner NSC component is its mass
within $200''$ up to where it dominates the star counts. This gives
$M_{\rm NSC}> 2 \times 10^7 M_\odot$.

Finally, we use our inferred dynamical cluster mass to update the
K-band mass-to-light ratio of the NSC. The best-determined mass is
within $100''$. Comparing our ${M_*}(r\!<\!100'')= (8.94\pm
0.31{|_{\rm stat}}\pm0.9{|_{\rm syst}})\times {10^6}{M_ \odot}$ with
the K-band luminosity of the old stars derived in \citet{fc2014}, $L_{100''}=(12.12\pm
2.58)\times 10^6 L_{\odot,{\rm Ks}}$, we obtain
$M/L_{Ks}=(0.76\pm0.18) M_\odot/L_{\odot,{\rm Ks}}$. The error is
dominated by the uncertainty in the luminosity (21\%, compared to a
total 10\% in mass from adding statistical and systematic errors in
quadrature).  The inferred range is consistent with values expected
for mostly old, solar metallicity populations with normal IMF
\citep[e.g.,][]{courteau2013,fc2014}.

\subsection{Evolution of the NSC} 

After $\sim 10$ half mass relaxation times $t_{rh}$ a dense nuclear
star cluster will eventually evolve to form a Bahcall-Wolf cusp with
slope $\gamma=7/4$ \citep{m2013}; for rotating dense star clusters
around black holes this was studied by \citet{fiestas2010}. The
minimum allowable inner slope for a spherical system with a black hole
to have a positive DF is $\gamma=0.5$.  From the data it appears that
the Galactic NSC instead has a core \citep{b2009,fc2014}, with the
number density possibly even decreasing very close to the center
($r<0.3$ pc). This is far from the expected Bahcall-Wolf cusp,
indicating that the NSC is not fully relaxed.  It is consistent with
the relaxation time of the NSC being of order $10$ Gyr everywhere in
the cluster \citep{m2013}.

From Fig.~\ref{plot_13} we see that the rotational properties of the
Milky Way's NSC are close to those of an isotropic rotator.
\citet{fiestas2012} found that relaxation in rotating clusters causes
a slow ($\sim 3 t_{rh}$) evolution of the rotation profile.
\citet{kim2008} found that it also drives the velocity dispersions
towards isotropy; in their initially already nearly isotropic models
this happens in $\sim 4 t_{rh}$. On a similar time-scale the cluster
becomes rounder \citep{e1999}.  Comparing with the NSC relaxation
time suggests that these processes are too slow to greatly modify the
dynamical structure of the NSC, and thus that its properties were
probably largely set up at the time of its formation.

The rotation-supported structure of the NSC could be due to the
rotation of the gas from which its stars formed, but it could also be
explained if the NSC formed from merging of globular clusters. In the
latter model, if the black hole is already present, the NSC density
and rotation after completion of the merging phase reflects the
distribution of disrupted material in the potential of the black-hole
\citep[e.g.][]{ac2012}. Subsequently, relaxation would lead to
shrinking of the core by a factor of $\sim 2$ in $10$ Gyr towards a
value similar to that observed \citep{m2010}. In the simulations of
\citet{ac2012}, the final relaxed model has an inner slope of
$\gamma=0.45$, not far from our models (note that in flattened
semi-isotropic models the minimum allowed slope for the density is
also 0.5 \citep{qh1995}). Their cluster also evolved towards a more
spherical shape, however, starting from a configuration with much less
rotation and flattening than we inferred here for the present Milky
Way NSC.  Similar models with a net rotation in the initial
distribution of globular clusters could lead to a final dynamical
structure more similar to the Milky Way NSC.

\subsection{Distance to the Galactic center}

From our large proper motion and los velocity datasets, we obtained
a new estimate for the statistical parallax distance to the NSC
using axisymmetric Jeans modeling based on the cluster's
inferred dynamical structure. From matching our best dynamical
model to the proper motion and los velocity dispersions within
approximately $\vert l\vert, \vert b \vert < 50''$, we found ${R_0} =
8.27 \pm 0.09{|_{\rm stat}} \pm0.1{|_{\rm syst}}$ kpc. The statistical
error is very small, reflecting the large number of fitted dispersion
points. The systematic modeling error was estimated from uncertainties
in the density structure of the NSC, as discussed in
Section~\ref{s_distance}.

Our new distance determination is much more accurate than that of
\citet{dm2013} based on anisotropic spherical Jeans models of the NSC,
$R_0=8.92_{-0.58}^{+0.58}$ kpc, but is consistent within their large
errors. We believe this is mostly due to the much larger radial range
we modeled, which leaves less freedom in the dynamical structure of
the model.

The new value for $R_0$ is in the range $R_0 = 8.33 \pm0.35$ kpc found
by \citet{ge2009} from analyzing stellar orbits around Sgr A$^\ast$.
A joint statistical analysis of the NSC data with the orbit results of
\citet{ge2009} gives a new best value and error ${R_0} = 8.33\pm0.11$
kpc (Fig.~\ref{plot_22}, Section~\ref{s_distance}). Our estimated
systematic error of $0.1$ kpc for $R_0$ in the NSC modeling translates
to a similar additional uncertainty in this combined $R_0$
measurement.

Measurements of $R_0$ prior to 2010 were reviewed by
\citet{geisen2010}.  Their weighted average of direct measurements is
$R_0=8.23\pm0.20\pm0.19$ kpc, where the first error is the variance of
the weighted mean and the second the unbiased weighted sample
variance. Two recent measurements give ${R_0} = 8.33 \pm
0.05{|_{\rm stat}} \pm0.14{|_{\rm syst}}$ kpc from RR Lyrae stars
\citep{dekany2013} and ${R_0} = 8.34 \pm 0.14$ kpc from fitting
axially symmetric disk models to trigonometric parallaxes of star
forming regions \citep{reid2014}.  These measurements are consistent
with each other and with our distance value from the statistical parallax
of the NSC, with or without including the results from stellar orbits
around Sgr A$^*$, and the total errors of all three measurements are
similar, $\sim 2\%$.

\subsection{Mass of the Galactic supermassive black hole}

Given a dynamical model, it is possible to constrain the mass of the
central black hole from 3D stellar kinematics of the NSC alone.  With
axisymmetric Jeans modeling we found ${M_ \bullet } = (3.86 \pm
0.14{|_{\rm stat}\pm 0.4{|_{\rm syst}}}) \times {10^6}{M_\odot}$,
where the systematic modeling error is estimated from the difference
between models with different inner cluster flattening as discussed in
Section~\ref{s_distance}. Within errors this result is in agreement
with the black hole mass determined from stellar orbits around Sgr
A$^*$ \citep{ge2009}.

Our dataset for the NSC is the largest analyzed so far, and the
axisymmetric dynamical model is the most accurate to date; it compares
well with the various proper motion and line-of-sight velocity
histograms. Nonetheless, future improvements may be possible if the
uncertainties in the star density distribution and kinematics within
20'' can be reduced, the effects of dust are incorporated, and
possible deviations from the assumed 2I-axisymmetric dynamical
structure are taken into account.

Several similar analyses have been previously made using spherical
isotropic or anisotropic modeling. \citet{tg2008} used isotropic
spherical Jeans modeling for proper motions and radial velocities in
$1''<R<100''$; their best estimate is $M_\bullet\sim 1.2 \times 10^6
{M_\odot}$, much lower than the value found from stellar
orbits. \citet{sm2009} constructed isotropic and anisotropic spherical
broken power-law models, resulting in a black hole mass of ${M_
  \bullet } = 3.6_{-0.4}^{+0.2}\times {10^6}{M_\odot }$.  However,
\citet{fc2014} find $M_\bullet\sim 2.27\pm 0.25 \times 10^6
{M_\odot}$, also using a power-law tracer density. They argue that the
main reason for the difference to \citet{sm2009} is because their
velocity dispersion data for $R>15''$ are more accurate, and their
sample is better cleaned for young stars in the central
$R<2.5''$. Assuming an isotropic spherical model with constant M/L,
\citet{fc2014} find $M_\bullet\sim 4.35\pm0.12 \times 10^6 {M_\odot}$.
\citet{dm2013} used 3D stellar kinematics within only the central
$0.5$ pc of the NSC.  Applying spherical Jeans modeling, they obtained
${M_\bullet } = 5.76_{-1.26}^{+1.76}\times{10^6}{M_ \odot }$ which is
consistent with that derived from stellar orbits inside $1''$, within
the large errors. However, in their modeling they used a very small
density slope for the NSC, of $\gamma = 0.05$, which does not
correspond to a positive DF for their quasi-isotropic model.

Based on this work and our own models in Section 4, the black hole
mass inferred from NSC dynamics is larger for constant M/L models than
for power law models, and it increases with the flattening of the
cluster density distribution.

The conceptually best method to determine the black hole mass is from
stellar orbits close to the black hole
\citep{schoe2002,ghez2008,ge2009}, as it requires only the assumption
of Keplerian orbits and is therefore least susceptible to systematic
errors. \citet{ge2009} find that the largest uncertainty in the value
obtained for $M_\bullet$ is due to the uncertainty in $R_0$, and that
$M_\bullet$ scales as $M_\bullet\propto R_0^{2.19}$. Therefore using
our improved statistical parallax for the NSC also leads to a more
accurate determination of the black hole mass. A joint statistical
analysis of the axisymmetric NSC modeling together with the orbit
modeling of \cite{ge2009} gives a new best value and error for the
black hole mass, ${M_ \bullet } = (4.26\pm 0.14) \times
{10^6}{M_\odot}$ (see Fig.~\ref{plot_22}, Section~\ref{s_distance}).
An additional systematic error of 0.1 kpc for $R_0$ in the NSC
modeling, through the BH mass-distance relation given in Gillessen et
al (2009), translates to an additional uncertainty $\simeq
0.1\times10^6 M_\odot$ in $M_\bullet$.

Combining this result with the mass modeling of the NSC, we can give
a revised value for the black hole influence radius $r_{\rm infl}$,
using a common definition of $r_{\rm infl}$ as the radius where the
interior mass $M(<r)$ of the NSC equals twice the black hole mass
\citep{m2013}. Comparing the interior mass profile in
Fig.~\ref{plot_8} as determined by the dynamical measurement with
$M_\bullet =4.26\times 10^6 M_\odot$, we obtain $r_{\rm infl}\simeq
94''=3.8$ pc.

The Milky Way is one of some 10 galaxies for which both the masses of
the black hole and of the NSC have been estimated
\citep{kormendy2013}. From these it is known that the ratio of both
masses varies widely. Based on the results above we estimate the Milky
Way mass ratio $M_\bullet/M_{\rm NSC}=0.12\pm0.04$, with the error
dominated by the uncertainty in the total NSC mass.

\section{Conclusions}

Our results can be summarized as follows:
\begin{itemize}

\item The density distribution of old stars in the central $1000''$ in
  the Galactic center can be well-approximated as the superposition of
  a spheroidal nuclear star cluster (NSC) with a scale length of $\sim
  100''$ and a much larger nuclear disk (NSD) component.

\item The difference between the proper motion dispersions $\sigma_l$
  and $\sigma_b$ cannot be explained by rotation alone, but is a
  consequence of the flattening of the NSC. The dynamically inferred
  axial ratio for the inner component is consistent with the axial
  ratio inferred from the star counts which for our two-component
  model is $q_1=0.73 \pm 0.04$.

\item The orbit structure of an axisymmetric two-integral DF
  $f(E,L_z)$ gives an excellent match to the observed double-peak in
  the $v_l$-proper motion velocity histograms, as well as to the
  shapes of the vertical $v_b$-proper motion histograms. Our model also
  compares well with the symmetrized (even) line-of-sight velocity
  histograms.

\item The rotation seen in the line-of-sight velocities can be
  modelled by adding an odd part of the DF, and this shows that the
  dynamical structure of the NSC is close to an isotropic rotator model.

\item Fitting proper motions and line-of-sight dispersions to the
  model determines the NSC mass within $100''$, the mass of the SMBH,
  and the distance to the NSC. From the star cluster data alone, we
  find ${M_*}(r\!<\!100'')\!=\!(8.94\!\pm\! 0.31{|_{\rm stat}}
  \!\pm\!0.9{|_{\rm syst}})\!\times\! {10^6}{M_\odot}$, ${M_\bullet }
  \!=\! (3.86\!\pm\!0.14{|_{\rm stat} \!\pm\! 0.4{|_{\rm syst}}})
  \!\times\! {10^6}{M_\odot }$, and ${R_0} \!=\! 8.27 \!\pm\!
  0.09{|_{\rm stat}}\!\pm\! 0.1{|_{\rm syst}}$ kpc, where the
  estimated systematic errors account for additional uncertainties in
  the dynamical modeling. The fiducial mass of the NSC is larger than
  in previous spherical models.  The total mass of the NSC is
  significantly more uncertain due to the surrounding nuclear disk; we
  estimate $M_{\rm NSC}\!=\!(2-6)\!\times\! 10^7 M_\odot$. The mass of
  the black hole determined with this approach is consistent with
  results from stellar orbits around Sgr A$^{*}$. The Galactic center
  distance agrees well with recent accurate determinations from RR
  Lyrae stars and masers in the Galactic disk, and has similarly small
  errors.
\item Combining our modeling results with the stellar orbit analysis
  of \citet{ge2009}, we find ${M_\bullet } \!=\!
  (4.23\!\pm\!0.14)\!\times\! {10^6}{M_\odot}$ and ${R_0} \!=\! 8.33
  \!\pm\! 0.11$ kpc. Because of the better constrained distance, the
  accuracy of the black hole mass is improved as well.  Combining with
  the parameters of the cluster, the black hole radius of influence is
  $3.8$ pc ($=94''$) and the ratio of black hole to cluster mass is
  estimated to be $0.12\!\pm\!0.04$.
\end{itemize}

\appendix

\section[]{Two-integral distributions functions}

In this part we give implementation instructions for the 2I-DF
algorithm of \citet[][HQ]{hq1993}. We will try to focus on the
important parts of the algorithm and also on the tests that one has to
make to ensure that the implementation works correctly. Our
implementation is based on \cite{qh1995} and made with \cite{w2011}.
For the theory the reader should consider the original HQ paper.

We will focus on the even part of the DF and for the case where the
potential at infinity, ${\Psi_\infty }$, is finite and therefore can
be set to zero. First one partitions the $(E,\eta)$ space where $\eta
\equiv {L_z}/{L_{z\,\max }}(E)$ takes values in $(0,1)$. The goal of
the HQ algorithm is to calculate the value of the DF on each of these
points on a 2D grid and subsequently end up with a 3D grid where we
can apply an interpolation to obtain the final smooth function
$f(E,{L_z})$. The energy values on the 2D grid are placed
logarithmically within an interval of interest $[{E_{\min }},{E_{\max
  }}]$ (higher $E_{\max }$ value is closer to the center) and the
values of $\eta$ are placed linearly between 0 and 1. Physically
allowable $E$ and $L_z$ correspond to bound orbits in the potential
$\Psi$ and therefore $E>0$. In addition at each energy there is a
maximum physically allowed $L_z$ corresponding to circular orbits with
$z=0$. This is given by the equations:

\begin{align}
\begin{array}{l}
E = \Psi (R_c^2,0) + R_c^2\frac{{d\Psi (R_c^2,0)}}{{d{R^2}}}{|_{R = {R_c}}}\\
L_z^2 =  - 2R_c^4\frac{{d\Psi (R_c^2,0)}}{{d{R^2}}}{|_{R = {R_c}}}
\label{A1}
\end{array}
\end{align}
where $R_c$ is the radius of the circular orbit and the value
${L_{z\,\max }} \equiv {L_z}({R_c})$ is the maximum allowed value of
$L_z$ at a specific $E$. The ${L_{z\,\max }}(E)$ function can be found
by solving the 1st equation for $R_c$ and substituting in the second
one therefore making a map $E \to {L_{z\,\max }}$. The value of the
potential of a circular orbit with energy $E$ is denoted by
${\Psi_{\rm env}}(E)$ and can be found from ${\Psi_{\rm env}}(E) =
\Psi (R_c^2,0)$ after solving the 1st of equation~(\ref{A1}) for
$R_c$. The value ${\Psi_{\rm env}}(E)$ is important for evaluation of
$f(E,{L_z})$ and it is used in the contour of the complex integral.

To calculate the even part $f_+(E,{L_z})$ of the DF for each point of
the grid we have to apply the following complex contour integral on
the complex $\xi$-plane using a suitable path:
\begin{align}
\begin{array}{l}
{f_+}(E,{L_z}) =\\
\frac{1}{{4{\pi^2}i\sqrt 2 }}\oint {\frac{{d\xi }}{{{{(\xi  - E)}^{1/2}}}}{\tilde\rho_{11}}\left( {\xi ,\frac{{L_z^2}}{{2{{(\xi  - E)}^{1/2}}}}} \right)}
\label{A2}
\end{array}
\end{align}
where the subscripts denotes the second partial derivative with
respect to the first argument. A possible path for the contour is
shown in figure \ref{A1}. The loop starts at the point 0 on the lower
side of the real $\xi$ axis, crosses the real $\xi$ axis at the point
${\Psi_{\rm env}}(E)$ and ends at the upper side of real $\xi$
axis. The parametrization of the path in general could be that of an
ellipse:

\begin{align}
\xi  = \frac{1}{2}{\Psi_{\rm env}}(E)(1 + \cos \theta) + ih\sin \theta ,\,\ - \pi  \le \theta  \le \pi
\label{A3}
\end{align}
where $h$ is the highest point of the ellipse. The value of $h$ should
not be too high because we want to avoid other singularities but not
too low either to maintain the accuracy. We optimize our
implementation by integrating along the upper part of the loop and
multiply the real part of the result by 2 (this is because of the
Schwarz reflection principle).

In order to calculate the integrand of the integral we need the following transformation:

\begin{align}
{\tilde\rho_{11}}(\xi ,{R^2}) = \frac{{{\rho_{22}}({R^2},{z^2})}}{{{{[{\Psi_2}({R^2},{z^2})]}^2}}} - \frac{{{\rho_2}({R^2},{z^2}){\Psi_{22}}({R^2},{z^2})}}{{{{[{\Psi_2}({R^2},{z^2})]}^3}}}
\label{A4}
\end{align}
in which each subscript denotes a partial differentiation with respect
to $z^2$. This equation is analogous to
equation~(\ref{transformation}) of the spherical case. In addition
${\tilde \rho}$ is the density considered as a function of $\xi$ and
$R^2$ as opposed to ${R^2}$ and ${z^2}$. The integrand of the contour
integral \ref{A2} depends only on $\theta$ angle for a given
$(E,{L_z})$ pair. Therefore we need the maps $R\to\xi$ and $z\to\xi$
in order to find the value of the integrand for a specific
$\theta$. The first map is given by
${R^2}=\frac{1}{2}L_z^2/(\xi-E)$. The second is given by solving the
equation ${\xi}=\Psi\left[ {\frac{{L_z^2}}{{2(\xi-E)}},{z^2}}\right]$
for $z$. It is very important that the solution of the previous
equation corresponds to the correct branch in which the integrand
attains its physically achieved values. In order to achieve that for
each pair $(E,{L_{z\,\max }})$ we start at the point $\xi={\Psi_{\rm
    env}}(E)(\theta=0)$ which belongs to the physical domain and we
look for the unique real positive solution. For the next point of the
contour we use as initial guess the value of $z$ from the previous
step that we already know that belongs to the correct branch. Using
this method we can calculate the integrand in several values of
$\theta$ then make an interpolation of the integrand and calculate the
value of the DF using numerical integration.

Figure \ref{plot_21} shows the shape of the DF for $\eta=0.5$ for the
potential we use in the fourth section of the paper for one value of
$h$, using the aforementioned procedure. We notice that for large
energies fluctuations of the DF appear. In order to solve this we
introduce a minor improvement of the procedure, by
generalizing the $h$ value of the contour to an energy-dependent
function $h = h(E)$.  The $h(E)$ could be a simple step function
that takes four or five different values. For our model the $h(E)$
function is a decreasing function of $E$. This means that the minor
axis of the ellipse should decrease as the $E$ increases to avoid such
fluctuations. In general we can write $h = h(E,L_z)$ so that the
contour depends both on $E$ and $L_z$.

Once we implement the algorithm it is necessary to test it. Our first
test is to check that the lower half of the integration path in figure
\ref{A1} is the complex conjugate of the upper half. Probably the next
most straightforward test is against the spherical case. It is
possible to use the HQ algorithm to calculate a DF for spherical
system. This DF should be equal to that obtained from Eddington's
formula for the same parameters. After calculating our 2I-DF we
compare its low-order moments with those of Jeans modeling. The 0th
and 2nd moments of the DF (the 1st is 0 for the even part) are given
from the integrals.

\begin{align}
\begin{array}{l}
\rho (R,z) = \frac{{4\pi }}{R}\int\limits_0^\Psi  {dE\int\limits_0^{R\sqrt {2(\Psi  - E)} } {d{L_z}{f_+}(E,{L_z})} } \\
\rho (R,z)\upsilon_\varphi^2(R,z) = \\
\frac{{4\pi }}{R}\int\limits_0^\Psi  {dE\int\limits_0^{R\sqrt {2(\Psi  - E)} } {d{L_z}{{\left( {\frac{{{L_z}}}{R}} \right)}^2}{f_+}(E,{L_z})} } \\
\rho (R,z)\upsilon_z^2(R,z) = \\
\frac{{2\pi }}{R}\int\limits_0^\Psi  {dE\int\limits_0^{R\sqrt {2(\Psi  - E)} } {d{L_z}\left[ {2(\Psi  - E) - {{\left( {\frac{{{L_z}}}{R}} \right)}^2}} \right]{f_+}(E,{L_z})} } 
\end{array}
\end{align}
Comparison with the 0th moment (density) is straight forward since the
density is analytically known from the start. The 1st moments should
be 0 within the expected error. In our implementation the error
between Jeans modeling and the DF is of the order of ${10^{ - 3}}$
within the area of interest. An additional test would be to integrate
the VPs over the velocity space.  Since the VPs integrals are
normalized with the surface density the integral of a VP over the
whole velocity space should be 1 within the expected error.

\begin{figure}
\centering
\includegraphics[width=\linewidth]{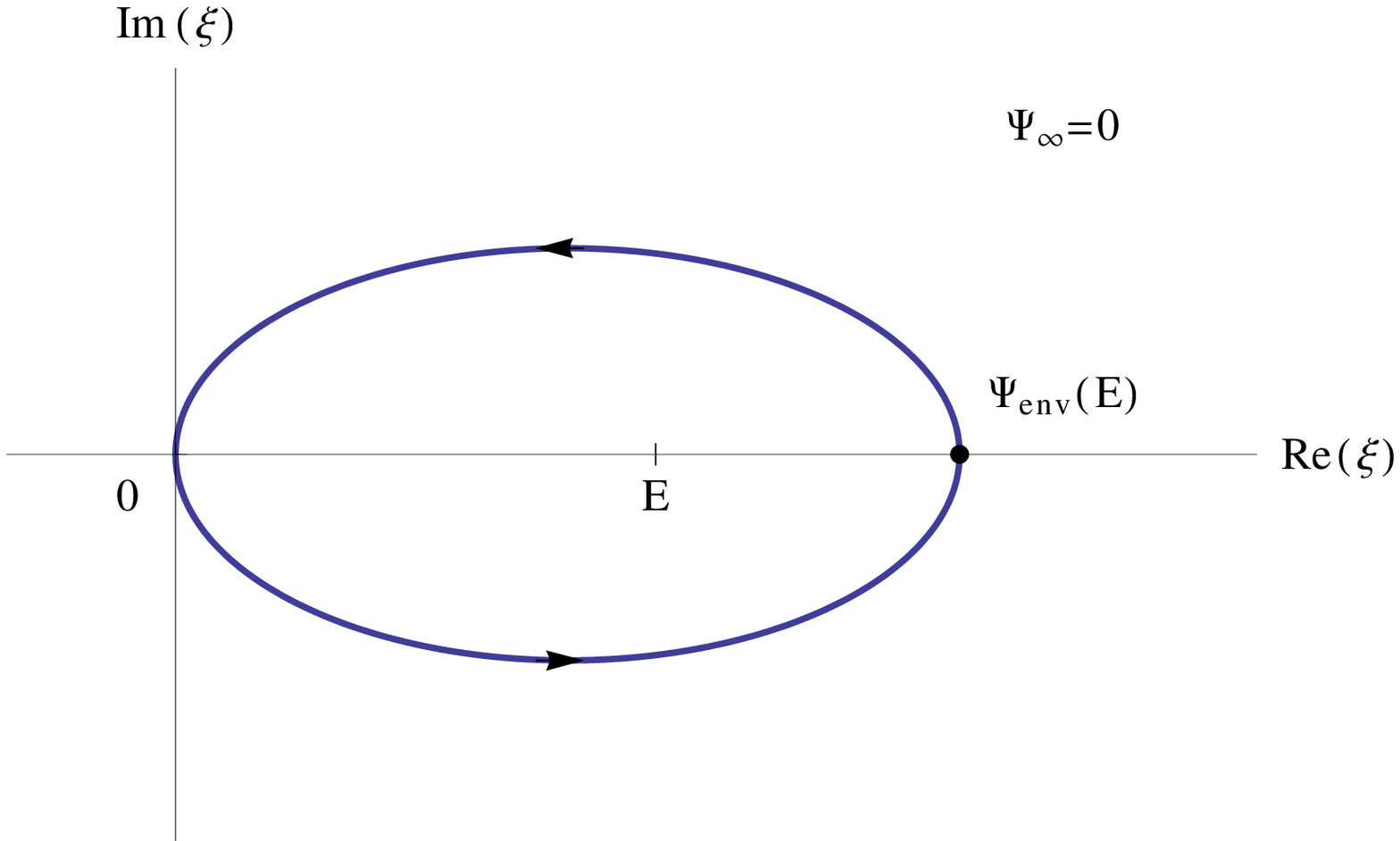}
\caption{The contour used for the numerical evaluation of $f(E,L_z)$
  for the case where ${\Psi_\infty }=0$. We optimize our
  implementation by integrating only along the upper or lower part and
  then multiplying the result by 2.}
\label{plot_20}
\end{figure}

\begin{figure}
\centering
\includegraphics[width=\linewidth]{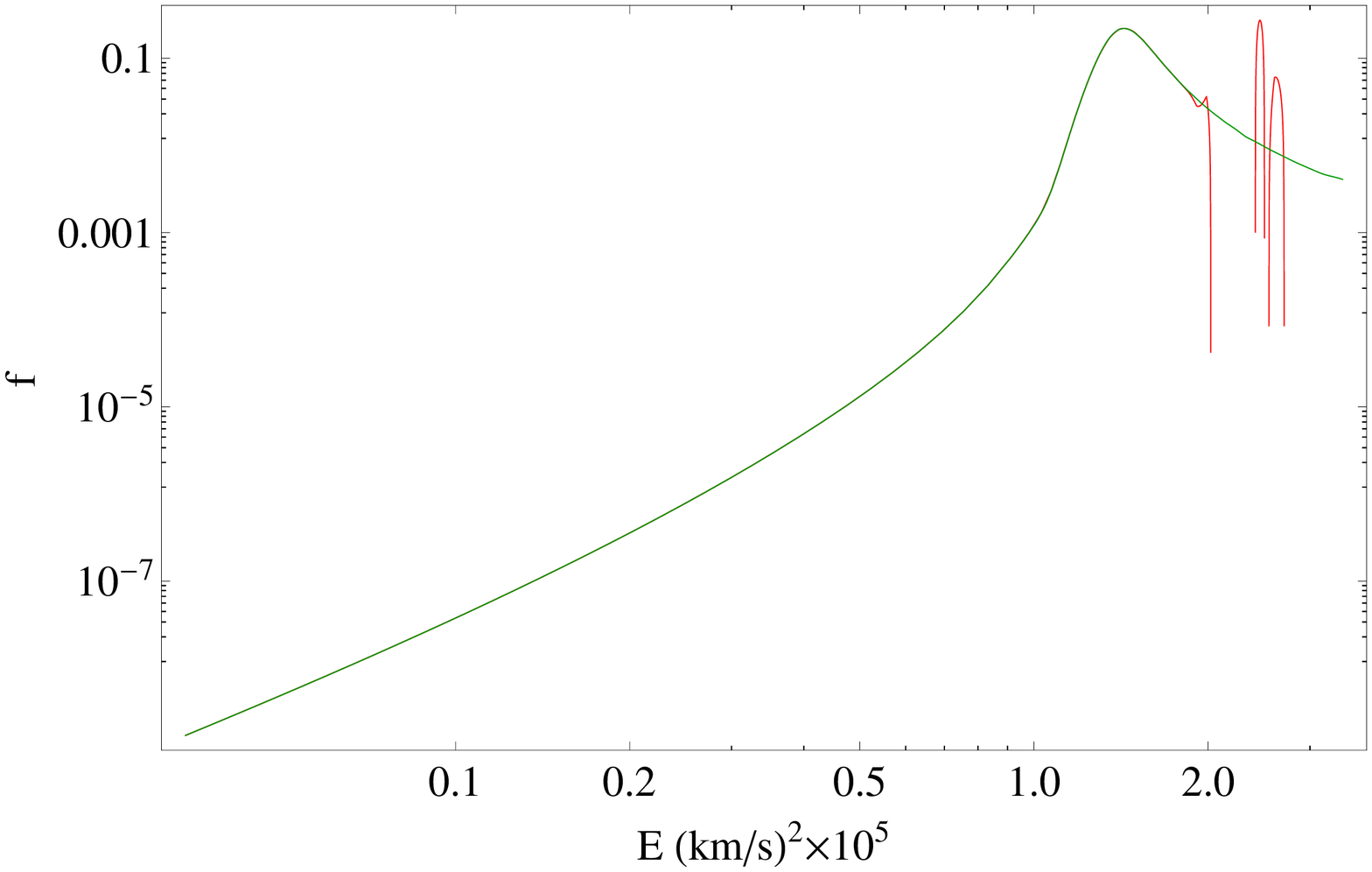}
\caption{This shows our best DF for $\eta=0.5$ (green
  line). Fluctuations (red lines) appear for large energies because we
  used a constant $h$ for equation~(\ref{A3}). To resolve this we used
  a more general function $h=h(E)$ or $h=h(E,L_z)$ even closer to the
  center.}
\label{plot_21}
\end{figure}

\FloatBarrier

\section[]{VELOCITY HISTOGRAMS FOR THE 2-I MODEL}

\begin{figure*}
\centering
\includegraphics[scale=0.5]{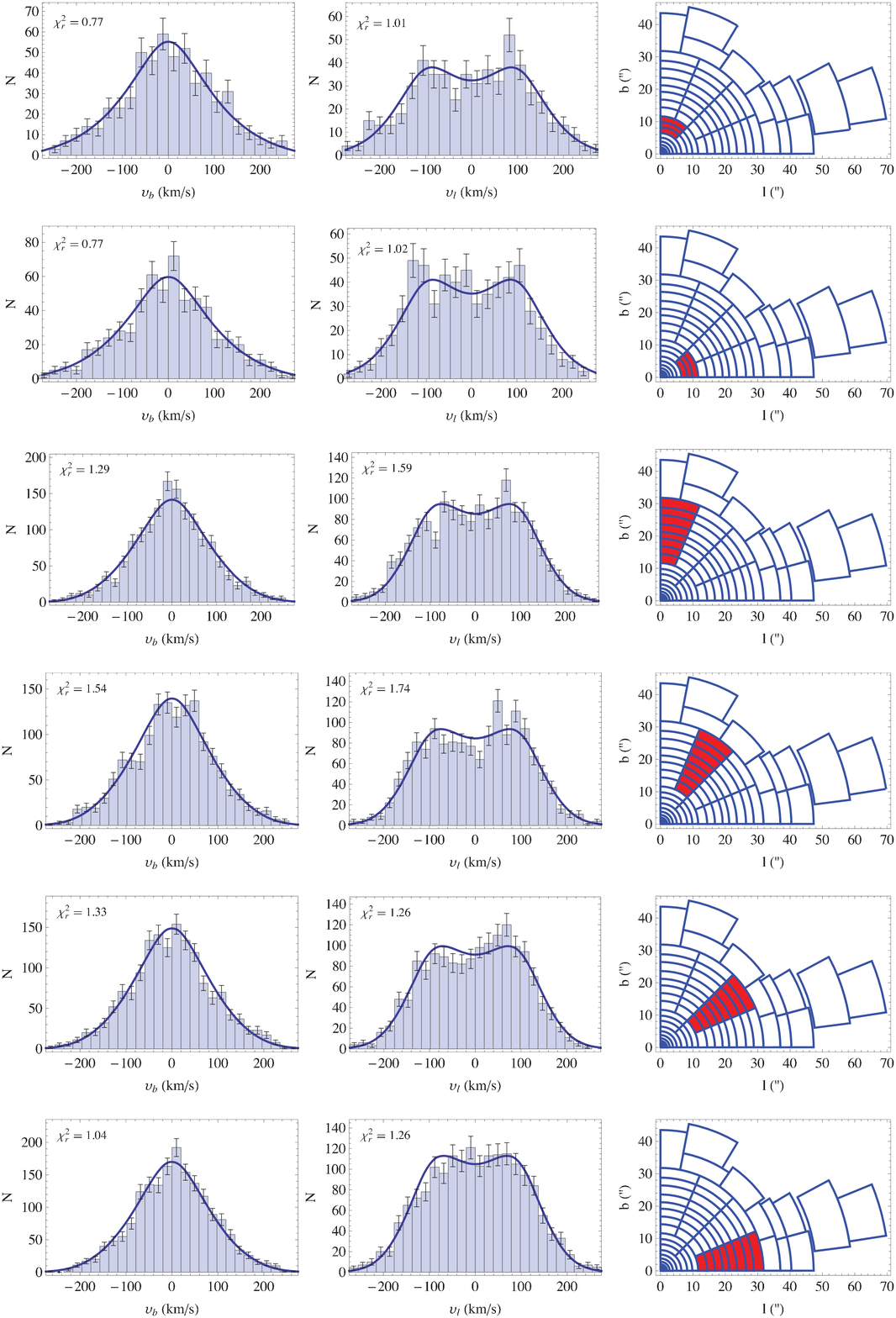}
\caption{VHs and VPs in the $l$ and $b$ directions predicted by the 2I
  model in angular bins. The reduced $\chi^2$ is also provided. The
  size of the bins is 0.6mas/yr ($\sim23.6$ km/s) for the upper two plots and 0.5mas/yr ($\sim19.6$ km/s)
  for the rest of the diagrams. The right column shows which cells
  have been used for the VHs and VPs.}
\label{plot_15}
\end{figure*}

\begin{figure*}
\centering
\includegraphics[width=\linewidth]{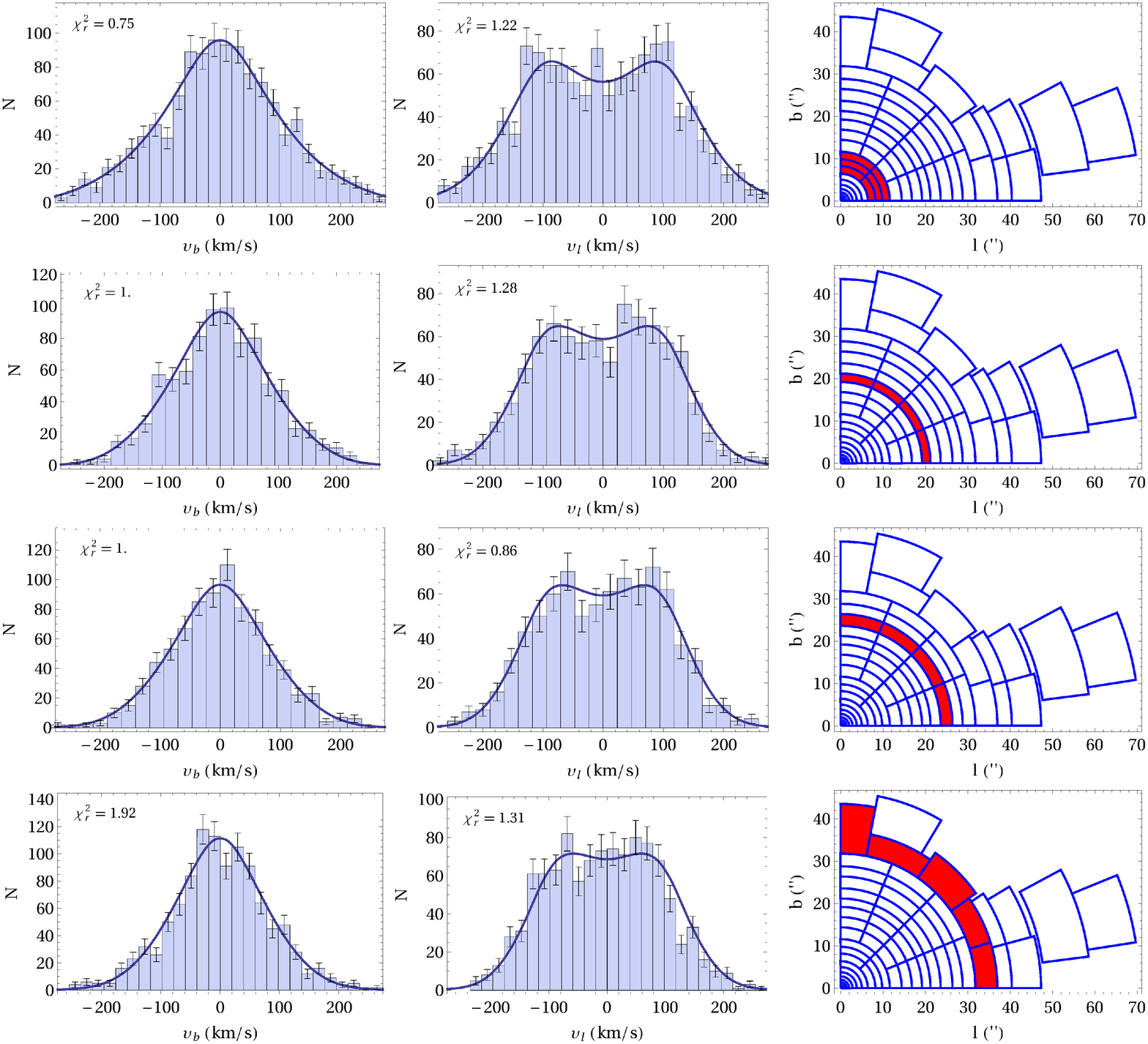}
\caption{VHs and VPs in the $l$ and $b$ directions predicted by the 2I
  model in radial bins. The reduced $\chi^2$ is also provided. The
  size of the bins is 0.5mas/yr ($\sim19.6$ km/s) for the 1st and 4th column and
  0.6mas/yr ($\sim23.6$ km/s) for the rest of the diagrams. The right column shows which
  cells have been used for the VHs and VPs.}
\label{plot_16}
\end{figure*}

\begin{figure*}
\centering
\includegraphics[width=\linewidth]{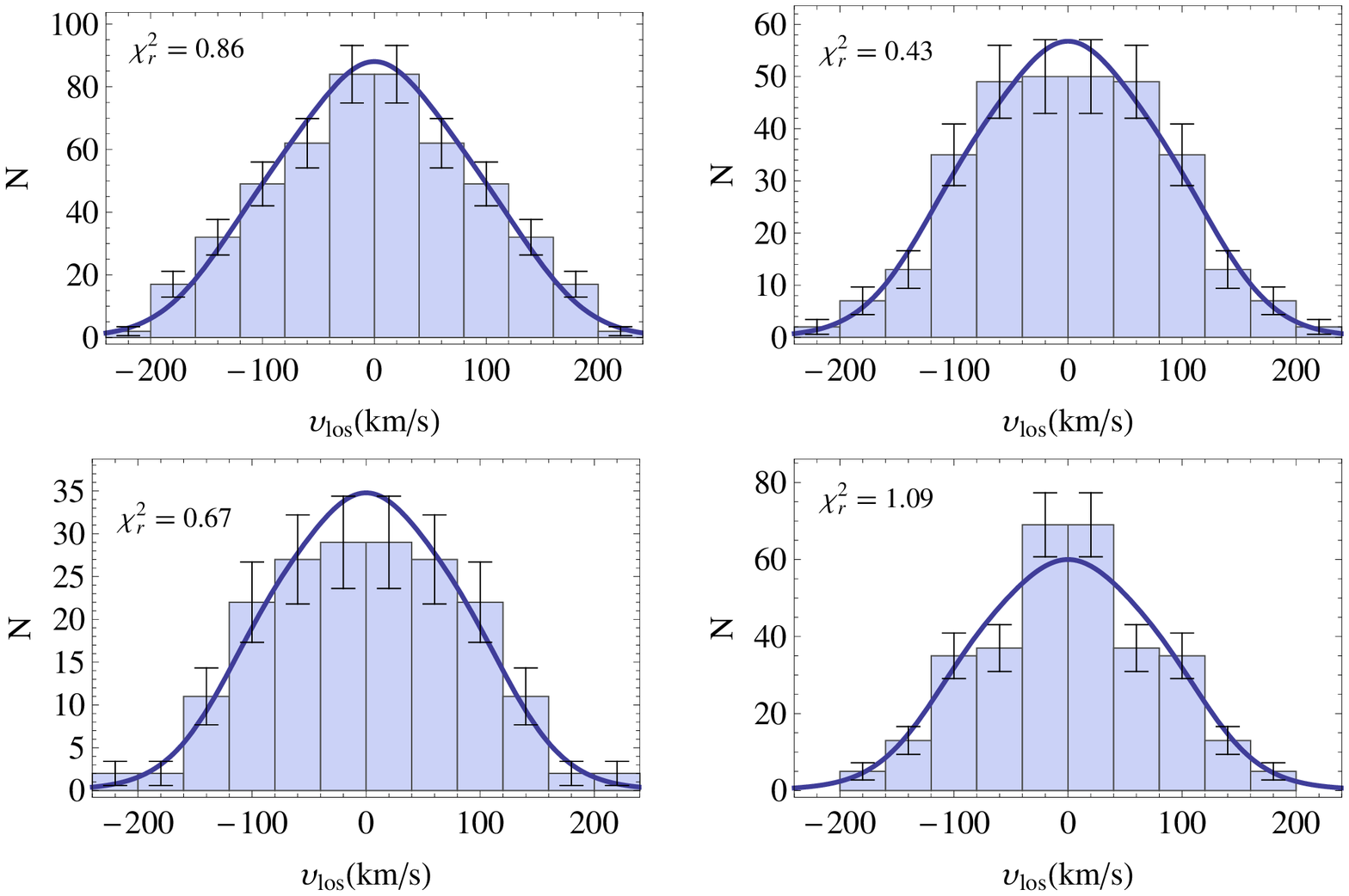}
\caption{VHs for the symmetrized los data compared with the
  corresponding even VPs of the model. The reduced $\chi^2$ is also
  provided. The size of the bins is 40km/s. For the upper left we use
  stars with $20''<|l|<30''$ and $|b|<20''$, for the upper right stars
  with $30''<|l|<40''$ and $|b|<20''$, for the bottom left
  $40''<|l|<50''$ and $|b|<20''$, and for the bottom right
  $50''<|l|<70''$ and $|b|<20''$.}
\label{plot_17}
\end{figure*}

\begin{figure*}
\centering
\includegraphics[width=\linewidth]{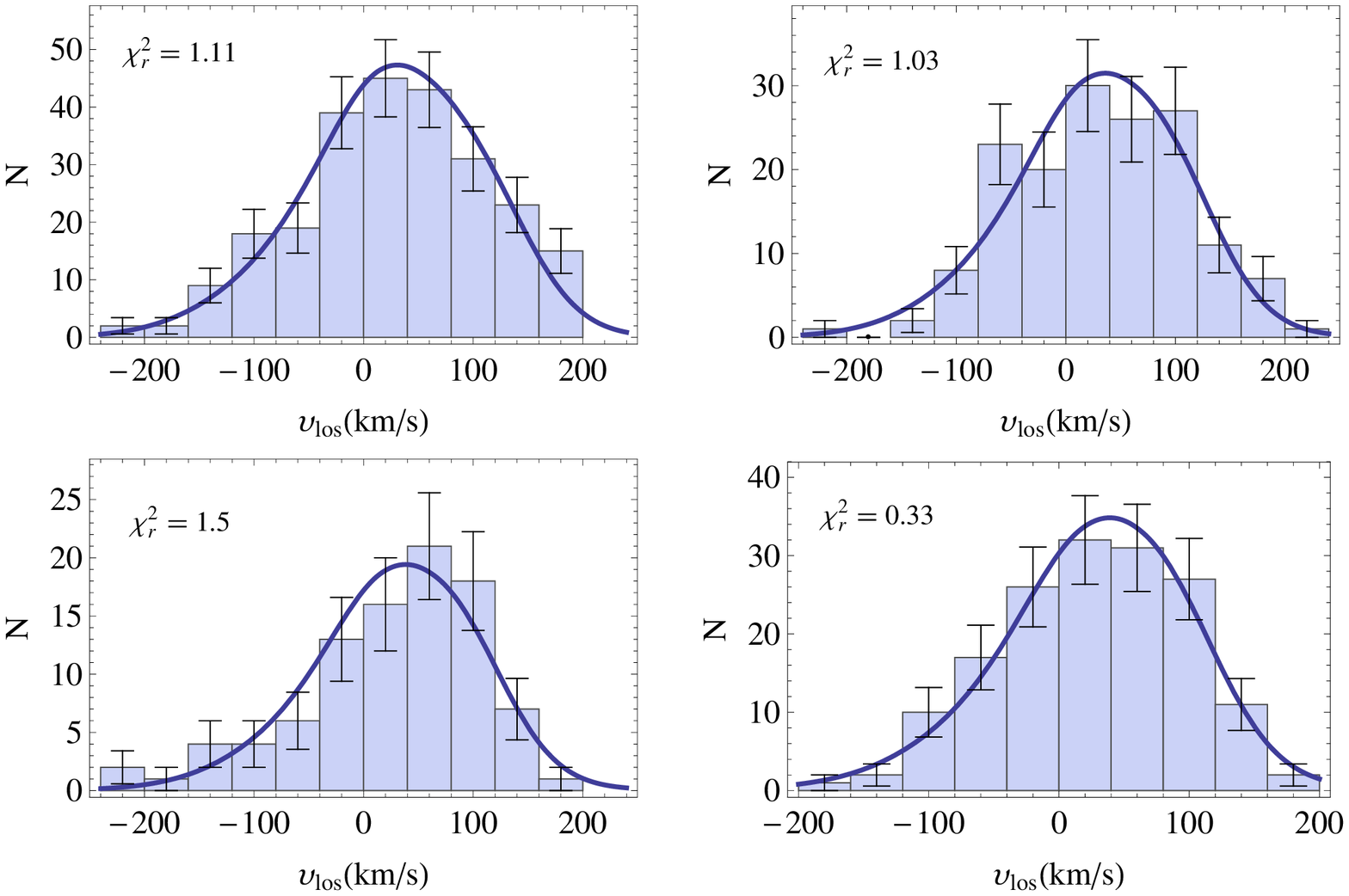}
\caption{Los VHs compared with the corresponding VPs of the model
  including rotation. The reduced $\chi^2$ is also provided. The size
  of the bins is 40km/s. For the upper left we use stars with
  $20''<|l|<30''$ and $|b|<20''$, for the upper right stars with
  $30''<|l|<40''$ and $|b|<20''$, for the bottom left $40''<|l|<50''$
  and $|b|<20''$, and for the bottom right $60''<|l|<80''$ and
  $|b|<20''$.}
\label{plot_19}
\end{figure*}

\label{lastpage}

\end{document}